\documentclass[useAMS,usenatbib, letterpaper]{mn2e}
\usepackage[totalwidth=500pt,totalheight=700pt]{geometry}
\usepackage{times}
\usepackage{graphicx}
\usepackage{epsfig}
\usepackage{ctable}
\usepackage{natbib}
\newcommand{\intl}{\textit{INTEGRAL}}
\newcommand{\rxte}{\textit{RXTE}}

\newcommand{\sw}{\textit{Swift}}

\def\cygx1{Cygnus~X$-$1}
\def\gx{GX~339$-$4}
\def\grs{GRS~1915+105}

\def\rsun{$R_{\odot}$}

\def\wm2{W~m$^{-2}$}
\def\mic{$\mu$m}
\def\cm2{cm$^{-2}$}
\def\se1{s$^{-1}$}
\def\nh{$N_{\rm H}$}
\def\nhe{N_{\rm H}}
\def\Av{$A_{\rm V}$}

\def\Ave{A_{\rm V}}

\usepackage{multirow}
\usepackage{array}
\usepackage{appendix}
\usepackage{booktabs}
\title[Optical and near-infrared spectroscopy of \gx]{Optical and
  near-infrared spectroscopy of the black hole \gx\\I. A focus on the continuum in the low/hard and
  high/soft states \thanks{Based on observations
  performed with European Southern Observatory (ESO) Telescopes at the
Paranal Observatory under programmes ID 284.D-5056 and 285.D-5007}}
\author[F. Rahoui et al.]{F. Rahoui$^{1,2}$\thanks{E-mail:
frahoui@cfa.harvard.edu}, M. Coriat$^{3}$, S. Corbel$^{4}$, 
M. Cadolle Bel$^5$, J.A. Tomsick$^6$, J.C. Lee$^{1,2}$,
\newauthor J. Rodriguez$^4$, D.M. Russell$^7$ and S. Migliari$^8$\\
$^{1}$Harvard University, Astronomy Department, 60 Garden street, Cambridge, MA 02138, USA\\
$^{2}$Harvard-Smithonian Center for Astrophysics, 60 Garden street, Cambridge, MA 02138, USA\\
$^{3}$School of Physics and Astronomy, University of Southampton, Southampton, Hampshire SO17 1BJ, UK\\
$^{4}$Laboratoire AIM (UMR 7158 CEA/DSM-CNRS-Universit\'e Paris
Diderot), Irfu/Service d'Astrophysique, CEA-Saclay, 91191, Gif-sur-Yvette Cedex, France\\
$^{5}$ESAC, ISOC, Villa\~nueva de la Ca\~nada, Madrid, Spain\\
$^{6}$Space Sciences Laboratory, 7 Gauss Way, University of California, Berkeley, CA 94720-7450, USA\\
$^{7}$Astronomical Institute Anton Pannekoek, University of Amsterdam, P.O. Box 94249, 1090 GE Amsterdam, The Netherlands\\
$^{8}$Departament d'Astronomia i Meteorologia, Universitat de Barcelona, Mart\'ii Franqu\`es 1, 08028 Barcelona, Spain}
\begin{document}

%\date{Accepted 1988 December 15. Received 1988 December 14; in original form 1988 October 11}

%\pagerange{\pageref{firstpage}--\pageref{lastpage}} \pubyear{2002}

\maketitle

\label{firstpage}

\begin{abstract}
The microquasar \gx, known to exhibit powerful compact jets
that dominate its radio to near-infrared emission, entered an
outburst in 2010 for the fifth time in about fifteen years. An
extensive radio to X-ray multi-wavelength campaign was
immediately triggered, and we report here on ESO/FORS2+ISAAC
optical and near-infrared spectroscopic observations, supported by ATCA
radio and {\it RXTE}/{\it Swift} X-ray quasi-simultaneous data. \gx\
was observed at three different epochs, once in the soft state
and twice in the hard state. In the soft state, the
optical and near-infrared continuum is largely consistent with the
Raleigh-Jeans tail of a thermal process. As an explanation, we favour irradiation of the
outer accretion disc by its inner regions, enhanced by disc
warping. An excess is also present at low frequencies, likely due to
a M subgiant companion star. During the first hard state, the optical/near-infrared
continuum is well-described by the optically thin synchrotron emission
of the compact jet combined with disc irradiation and perhaps another
component peaking in the ultraviolet. The spectral break where the
jet transits from the optically thick to thin regimes, located below
$1.20\times10^{14}$~Hz, is not detected and the extension of the optically thin
synchrotron is consistent with the 3--50~keV spectrum. In contrast, the emission during the second
hard state is more difficult to understand and points toward
a more complex jet continuum. 
In both cases, the near-infrared continuum is found to be variable at
timescales at least as short as 20~s, although these variabilities are
smoothed out beyond a few hundred seconds. This implies rapid variations -- in
flux and frequency -- of the location of the spectral break, i.e.
dramatic short  timescale changes of the physical conditions at the
base of the jet, such as the magnetic field and/or the base radius. 
\end{abstract}


\section{Introduction}
\begin{table*}
\caption{\small Summary of all the observations of \gx\ we made
  use of in this study. We give the observation number, the day of
  observation with FORS2 and ISAAC (in MJD), the day of quasi-simultaneous
  coverage with \rxte, \sw, and the ATCA telescope, as well as the ATCA
  flux levels at 5.5 and 9.0~GHz (in mJy).\label{logobs}}
\begin{tabular}{cccccccc}
\hline
Obs.~\#&FORS2&ISAAC&\rxte&\sw&ATCA&\multicolumn{2}{c}{ATCA flux density}\\
&&&&&&5.5~GHz&9.0~GHz\\
\hline
\hline
1&55261.39&55261.37&55261.19&55261.05&55261.89&9.02$\pm$0.10&9.60$\pm$0.05\\
2&55262.39&55262.38&55262.93&--&55262.91&8.24$\pm$0.05&8.05$\pm$0.10\\
3&55307.29&55307.24&55307.02&--&$-$&$-$&$-$\\
\hline
\end{tabular} 
\end{table*}

While microquasars radiate over the whole electromagnetic spectrum,
the optical and near-infrared (near-IR) domains are of
strong interest. Indeed, most of the components in the system can be detected
at these wavelengths, including the companion star, the accretion
disc, and material ejecta. In the last category, the best known are the
so-called compact jets, which are only detected in the hard state (HS)
and emit both through optically thick and thin synchrotron radiation. Within
the most accepted scheme \citep{1979Blandford, 1995Falcke}, their
emission is well-modelled by a flat or weakly-inverted power law (optically thick regime,
$F_\nu\propto\nu^\alpha$ with $-0.1\le \alpha \le 0.7$) from the radio to
some spectral break, beyond which $\alpha$ ranges between $-0.4$ and $-1$
depending on the electron energy distribution (optically thin
regime). The location of the break, thought to occur in the IR 
domain, is a crucial piece of information as it is closely related to the
physical conditions at the base of the jet, such as the magnetic field, the base
radius of the jet, and the total energy of the electron population. So
far, it has only been detected in three sources, \gx, 4U~0614+091, and
\cygx1\ \citep{2002Corbel, 2011Gandhi, 2006Migliari,
  2010Migliari, 2011Rahoui}, with additional observational evidence that
this break can move with luminosity \citep{2009Coriat, 2011Rahoui, 2011Gandhi, 2011Russell}. 

If a broken power law is a good approximation of a compact jet,
this very simple modelling is also plagued with severe
limitations. First, it is doubtful that the jet can sustain a
single power law-like spectrum from the radio to the spectral break,
and a change of the optically thick spectral index is
expected. Moreover, their emission could simply be more complicated
than a broken power law, in particular when cooling is taken into
account \citep{2009Peer}. Finally, compact jets could contribute to the
optical/near-IR (OIR) through other processes such as pre-shock synchrotron,
which would enhance significantly their emission at these wavelengths
\citep[see e.g.][]{2005Markoff}. OIR
spectroscopic observations of microquasars are therefore
crucial for correctly constraining the properties of compact jets;
\gx\ is obviously one of the best targets to fulfil this
goal as the jets strongly dominate the OIR
emission in the HS \citep[see e.g.][]{2004Buxton,2005Homanb}.

In early 2010, \gx\ entered an outburst, only the fifth time in
fifteen years. Considering the rarity of such an event, we triggered a
large multi-wavelength observational campaign of the source, which
included X-ray (\intl, \rxte, \sw), OIR (ESO, Faulkes), and
radio (ATCA). Here, we report on the OIR spectroscopy of
the source in the HS and soft state (SS), focusing on the spectral
continuum in the presence and absence of compact jets. A companion paper
\citep{2011Cadolle} presents an overview of this multi-wavelength
campaign, with a particular focus on X-ray behaviour. Finally,
a detailed study of the spectroscopic content of
the OIR spectra will be object of a third paper
(Rahoui et al., in prep). Sect.~2 is devoted to the
data reduction and analysis. In Sect.~3, we present spectral
energy distribution (SED) modelling. We discuss the outcomes in
Sect.~4 and conclude in Sect.~5.

\section{Observations and data reduction}

\gx\ was observed through low-resolution spectroscopy on 2010 March 6
(hereafter Obs.~1), March 7 (Obs.~2), and April 21 (Obs.~3)
quasi-simultaneously in the optical with ESO/FORS2 and
in the near-IR with ESO/ISAAC, mounted on the VLT/UT2 and UT3,
respectively  (PI F. Rahoui). In the framework of the Galactic bulge monitoring, the source was also 
quasi-simultaneously observed with the Rossi X-ray Timing Experiment
(\rxte, the data are immediately public). Moreover,
the \sw\ satellite also observed the field of \gx\ on a regular basis during the 2010 outburst, and one of
these pointed observations (Obs. ID 00030943010, P.I. M. Cadolle Bel) was conducted quasi-simultaneously to Obs.~1. 
Finally, \gx\ was detected in the radio domain during Obs.~1
and 2 with the Australia Telescope Compact Array (ATCA, P.I. S. Corbel)
at 5.5 and 9.0~GHz. All these observations are listed in Table~\ref{logobs}.

\subsection{Near-IR ESO/ISAAC observations}
\begin{table*}
\renewcommand{\arraystretch}{1.3}
\caption{\small  Best parameters obtained from the fits to the
   \gx\ X-ray spectra during Obs.~1, 2, and 3, with the model {\sc
   phabs$\times$reflect$\times$(diskbb+gaussian+powerlaw)}. The
 errorbars are given at the 90\% confidence level.\label{parobsx}}
\begin{tabular}{cccccccccc}
\hline
\hline
Obs.~\#&$\nhe$&$\Omega/2\pi$&$T_{\rm in}$&$N$&$E_{\rm iron}$&$\sigma_{\rm
  iron}$&$\Gamma$&--&$\chi_{\rm r}^2$ (dof)\\
--&{\tiny ($\times10^{22}\,{\rm cm}^{-2}$)}&&{\tiny (keV)}&{\tiny
  $\left(\frac{R_{\rm in}/{\rm km}}{D/{\rm 10\,kpc}}\right)^2 \cos(i)$}&{\tiny (keV)}&{\tiny
  (keV)}&{\tiny --}&{\tiny --}&{\tiny --}\\
\hline
1&$0.58_{-0.05}^{+0.05}$&$0.49_{-0.09}^{+0.10}$&$0.23_{-0.02}^{+0.03}$&$3.50_{-2.18}^{+4.44}\times10^4$&$6.31_{-0.23}^{+0.19}$&$0.81_{-0.30}^{+0.35}$&$1.73_{-0.02}^{+0.02}$&--&1.06 (651)\\
2&0.58 (fixed)&$0.39_{-0.09}^{+0.11}$&$-$&$-$&$6.24_{-0.22}^{+0.18}$&$0.62_{-0.31}^{+0.34}$&$1.72_{-0.03}^{+0.03}$&--&0.95 (74)\\
3&0.58 (fixed)&$-$&$0.83_{-0.04}^{+0.02}$&$3.03_{-0.42}^{+0.88}\times10^3$&$5.97^{+0.44}$&$1.33_{-0.31}^{+0.48}$&$2.46_{-0.16}^{+0.15}$&-&1.03 (73)\\
\hline
\end{tabular}
\end{table*}

\gx\ was observed through the SW$-$LR1 arm (1\arcsec slit
aperture) with the {\it J}, {\it H}, and {\it K} filters ($R\sim600$
and 1.10$-$2.5~\mic\ spectral coverage). The sky was clear,
and the airmass and seeing at corresponding central wavelengths were lower
than 1.2 and 0\farcs75, respectively. The total exposure time in each
filter was 240~s, divided in $4\times3\times20$~s dithered frames for
standard ABBA sky removal. The same nights, the telluric standard
stars HIP~84086 (B8/9V, March 6 2010), HIP~86735 (B5/6III, March 7
2010), and HIP~68124 (B2/3V, April 21 2010),  were
observed under the same conditions.

We reduced all the data with {\tt IRAF}\footnote{{\tt IRAF} is
    distributed by the National Optical Astronomy Observatories, which
    are operated by the Association of Universities for Research in
    Astronomy, Inc., under cooperative agreement with the National
    Science Foundation.} by performing correction for crosstalk and
bad pixels, flatfielding, and sky subtraction. The spectra
were then extracted, wavelength-calibrated by
matching the spectral dispersion axis with that of an ArXe arc, and
the atmospheric signatures were corrected with the \texttt{telluric}
task. This routine aligns the source's and standard star's spectra and scale
them according to their respective
airmass; they are then divided to remove the telluric features and
atmospheric extinction. The output, once corrected for
the exposure time discrepancies between the source and the standard
star observations, can therefore be multiplied by the
flux-calibrated standard star's spectrum to obtain that of the
source. Unfortunately, very few spectro-photometric standard stars
exist in the near-IR, and none of them was observed during our
program. The way we proceeded to flux-calibrate the \gx's ISAAC
spectra therefore deserves some detailed explanations.

We started from three synthetic Kurucz spectra corresponding to
the HIP~84086, HIP~86735, and HIP~68124 spectral types. They were then
reddened, convolved by Cousins-Bessel and 2MASS filters, and scaled to
match, through $\chi^2$ fitting, the measured {\it B}, {\it V}, {\it J}, {\it
    H}, and {\it K} flux densities of our three standard
  stars (see Figure~\ref{stdtheo}). The initial value of the optical extinction
\Av\ along their line-of-sight (LOS) were obtained from their colour
indices using the standard relation $\Ave=R_{\rm V}\times E(B-V)$ with
$R_{\rm V}=3.1$ (corresponding to the diffuse ISM) and
$E(B-V)=(B-V)-(B-V)_0$. The intrinsic colours $(B-V)_0$ were taken
from \citet{1984Cramer} for the corresponding spectral types, and we
made use of the extinction law given in \citet{1999Fitzpatrick} to get the
$A_{\lambda}$ values at other wavelengths. Once flux-calibrated, the
standard stars' synthetic spectra were
multiplied by the \texttt{telluric} outputs to get the final \gx's
flux-calibrated spectra. It is worth mentioning that the ESO/REM
\textit{J}, {\it H}, and \textit{K} photometric observations
quasi-simultaneous with Obs.~1 \citep{2011Cadolle}, are
consistent, with less than 5\% discrepancies, with our
result. Therefore, in our analysis, we adopt an intrinsic 5\% uncertainty
for the flux-calibration. 

\subsection{Optical ESO/FORS2 observations}

We used the 300V and 300I grisms
(1\arcsec\ aperture) with the GG435 and OG590 filters,
respectively. The exposure time was set to
120~s in each filter, and the atmospheric conditions were
good, with a clear sky and airmass and seeing lower than 1.14 and
0\farcs7, respectively.  For flux calibration purposes, the ESO
spectro-photometric standard stars Hiltner~600 (B3V) and
CD$-$32~9927 (A0V) were observed under the same
conditions. We reduced the data using the {\tt IRAF} suite. The
process consisted of bad pixels correction, bias subtraction,
flatfielding, spectral extraction, and wavelength calibration by
matching the spectral dispersion axis with that of an HeAr arc lamp. The
resulting \gx\ 1D spectra were then flux-calibrated using the
CTIO extinction and flux tables\footnote{{\it ctioextinct.dat}, {\it h600.dat}, and
{\it cd32.dat}} available with {\tt IRAF}. The flux levels in the
overlapping region of the GG435 and OG590 filters appear consistent, 
with discrepancies less than 5\%, which allowed us to safely combine
them.  However, the strong limitation of this method is
that the resulting optical spectra do not overlap with the near-IR
ones because the ESO spectro-photometric standard stars are not defined
beyond 1.02~\mic. To cover the gap between the
optical and near-IR, we also reduced and flux-calibrated the FORS2
spectra with the same procedure as that used for the ISAAC data (see
Figure~\ref{stdtheo2}, for the flux-calibrated synthetic
Kurucz spectra). This gives very similar results, as showed in
Figure~\ref{compopt} (left panel). The optical spectra flux-calibrated with the ESO
spectro-photometric standards are displayed in green, those
flux-calibrated with synthetic spectra in red. We also display the
{\it J} band spectra in blue, and a good match between the
optical and near-IR continuum levels is reached (right
  panel). Note that the displayed red spectra were rebinned to a third of their
  resolution to improve the visibility of the optical/\textit{J} overlapping
  region. However, only the original ones were used for SED fitting.

\subsection{\rxte\ and \sw\ observations}

We present here the reduction and analysis of the high-energy data we
made use of. We use this only for the determination of the spectral state of \gx\ during
the FORS2+ISAAC observations, as well as the level of emission from
the main components, i.e. the accretion disc and the corona/jet. 
A far more comprehensive X-ray study of \gx\ during the 2010
outburst is presented in \citet{2011Cadolle}.

\subsubsection{Data reduction}
\begin{figure*}
\begin{center}
\begin{tabular}{ccc}
$K$&$H$&$J$\\
\hline
\hline
&&\\
&HS1&\\
\includegraphics[width=5.6cm]{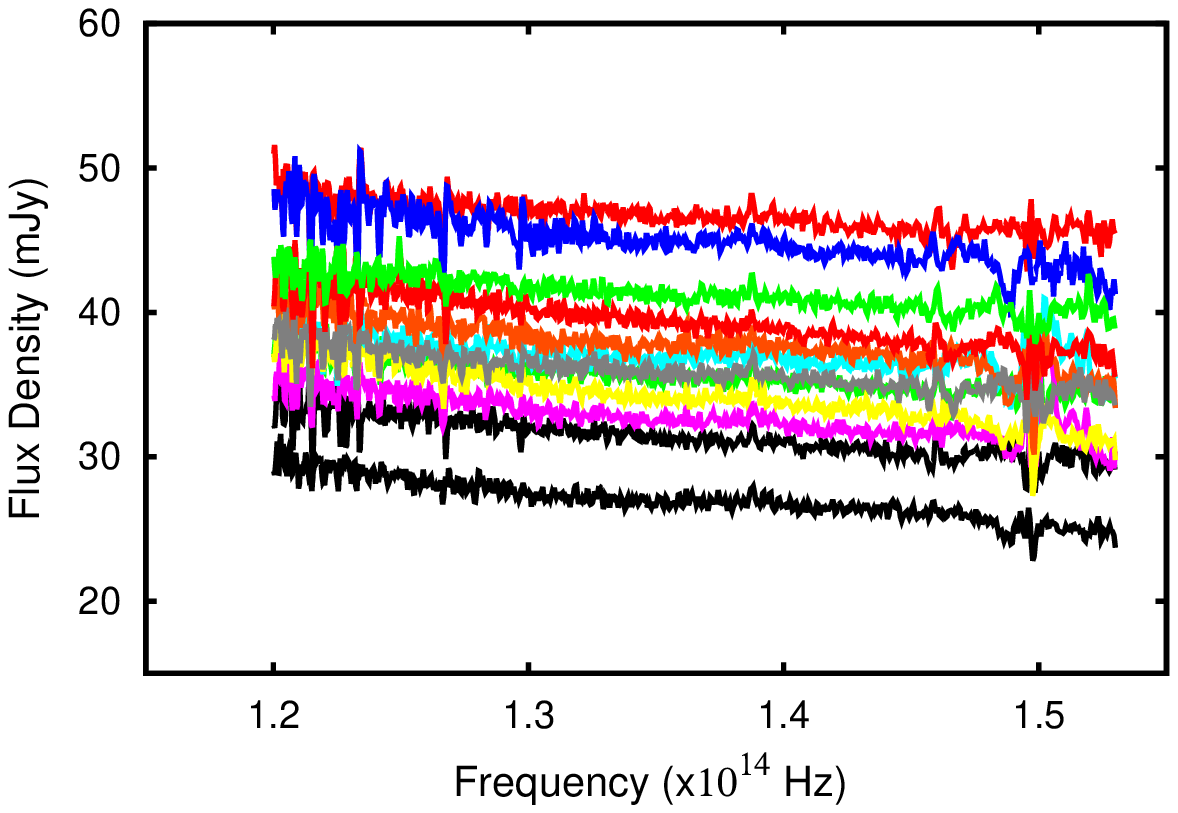}&\includegraphics[width=5.6cm]{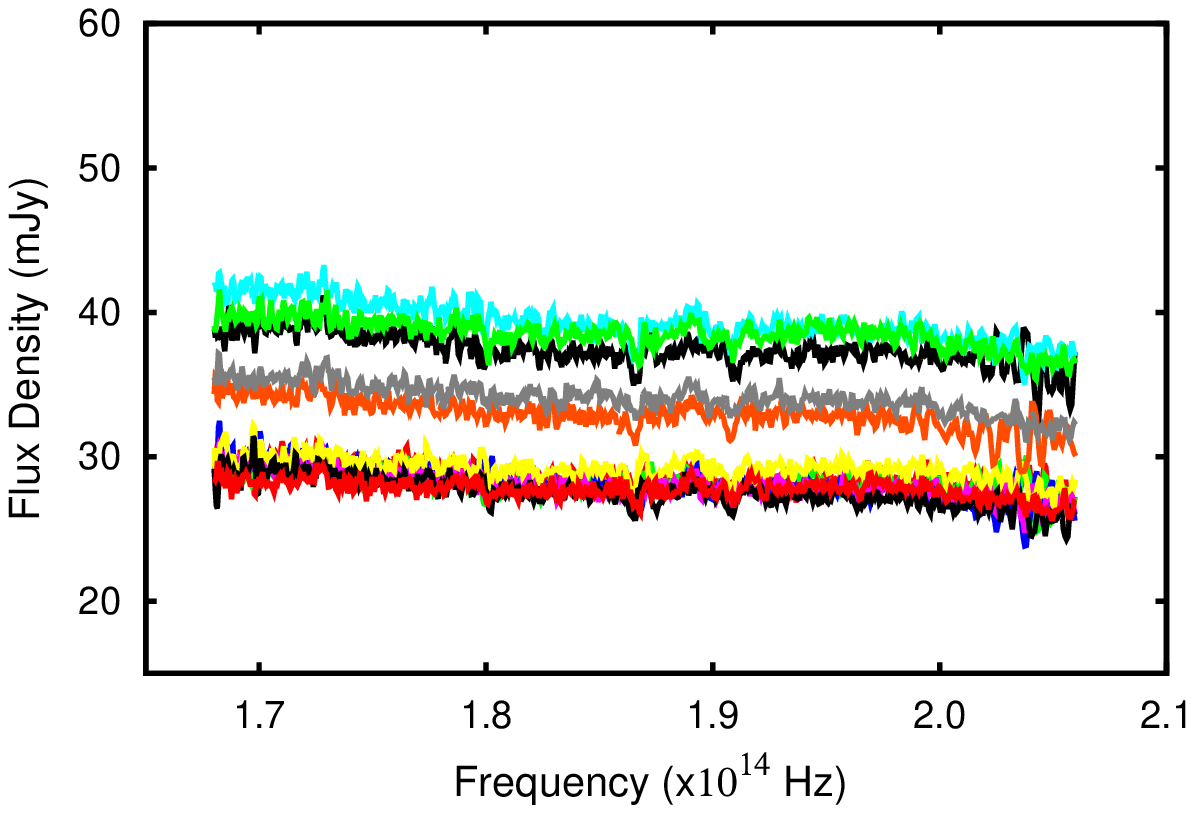}&\includegraphics[width=5.6cm]{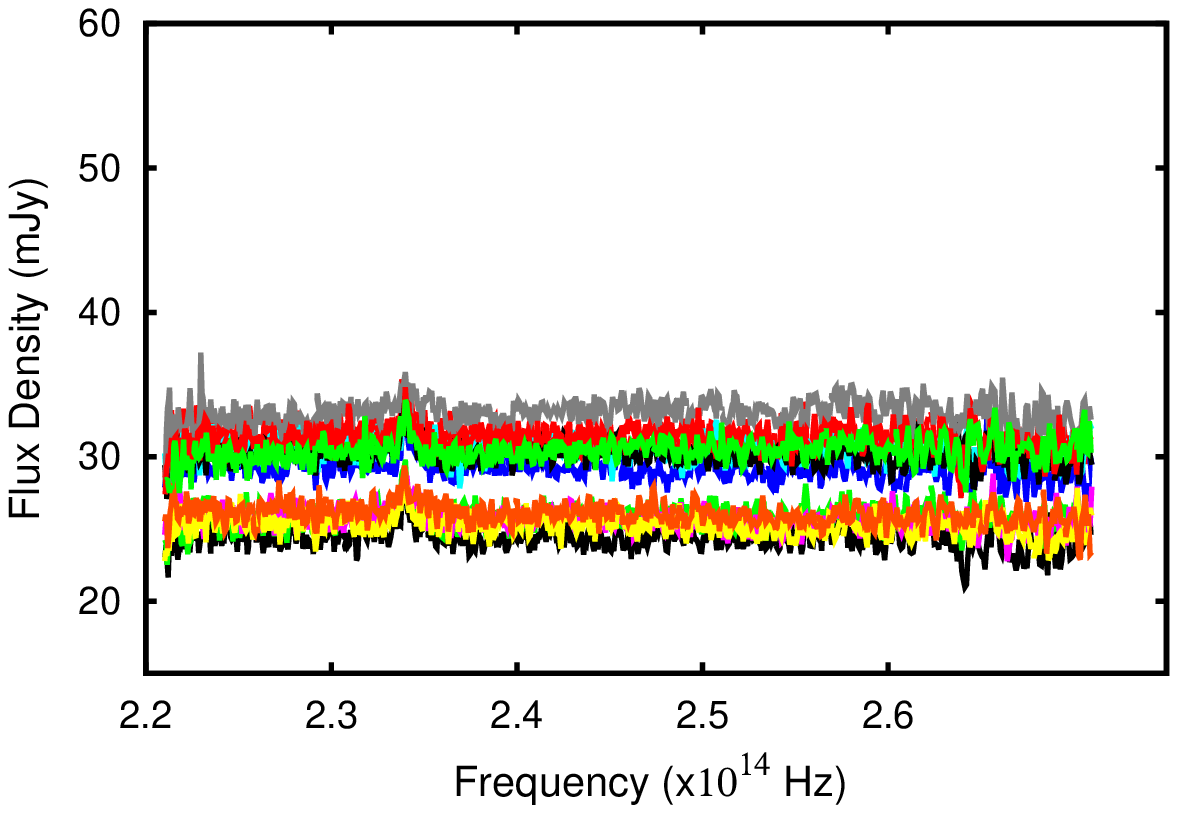}\\
&HS2&\\
\includegraphics[width=5.6cm]{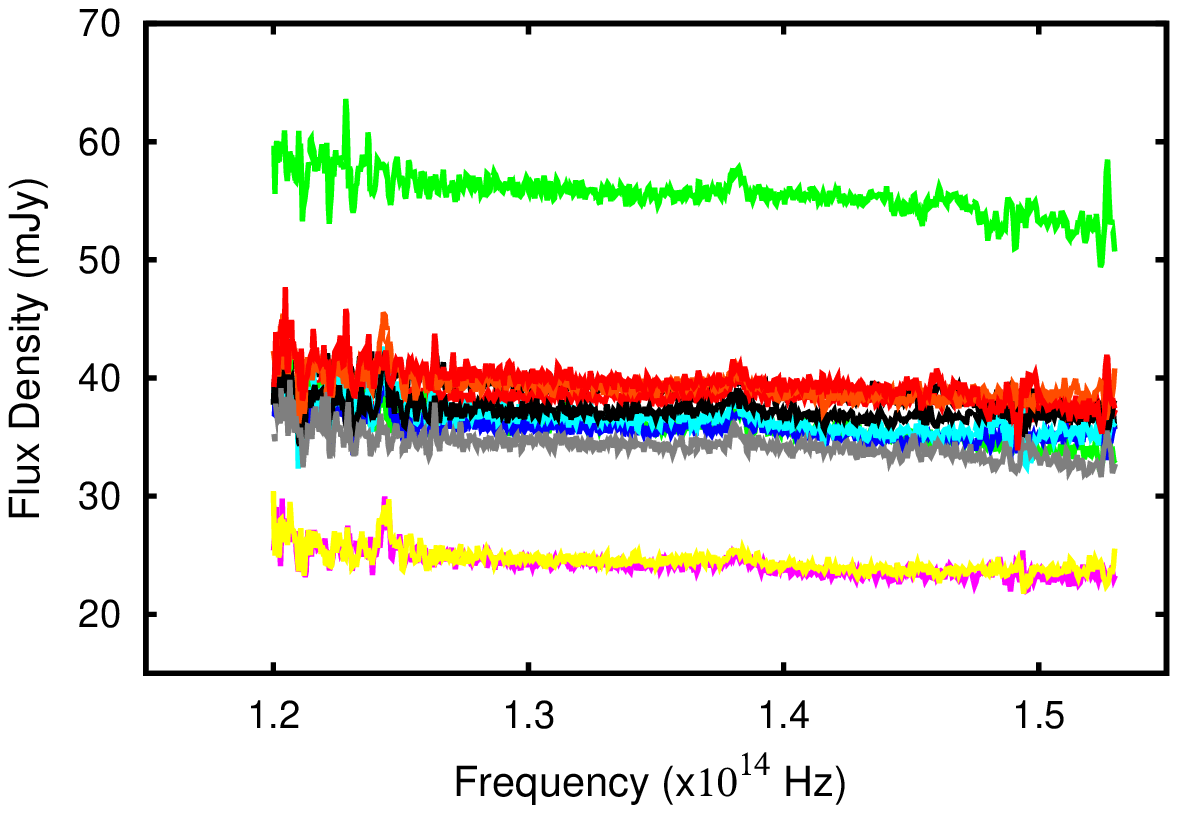}&\includegraphics[width=5.6cm]{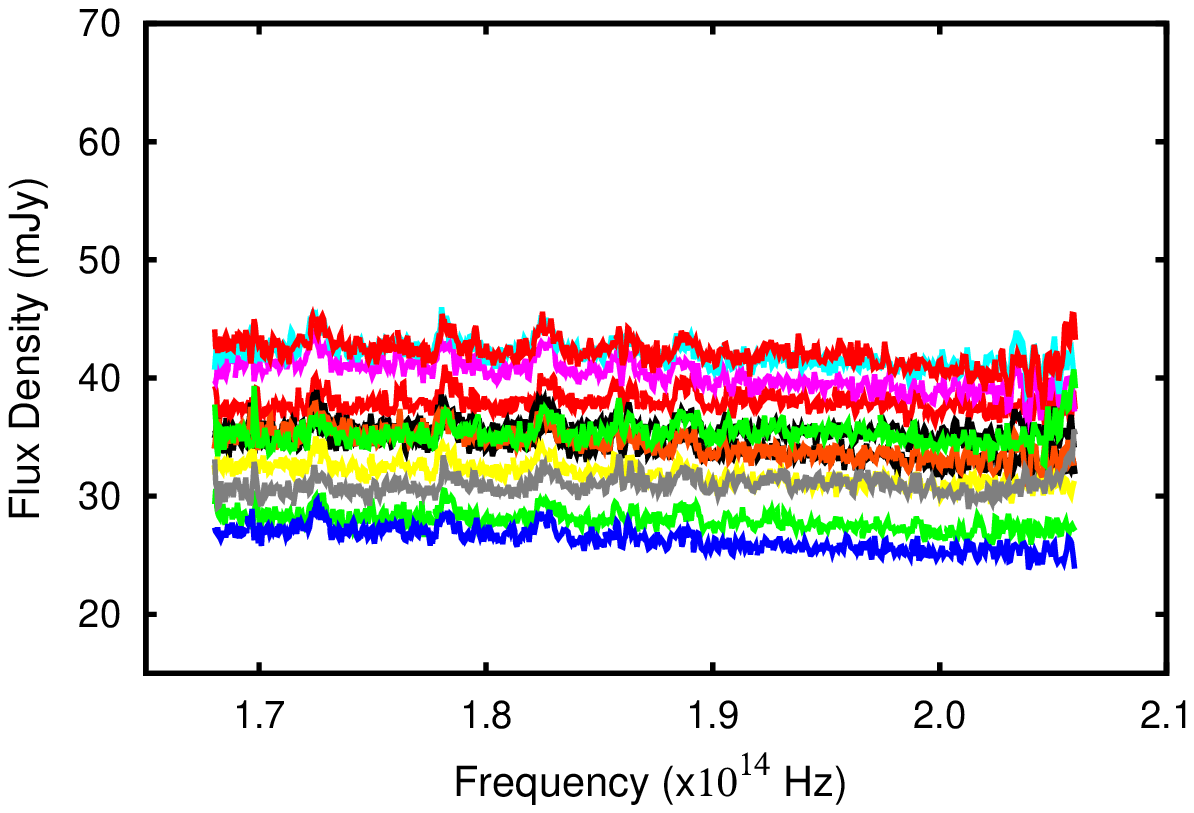}&\includegraphics[width=5.6cm]{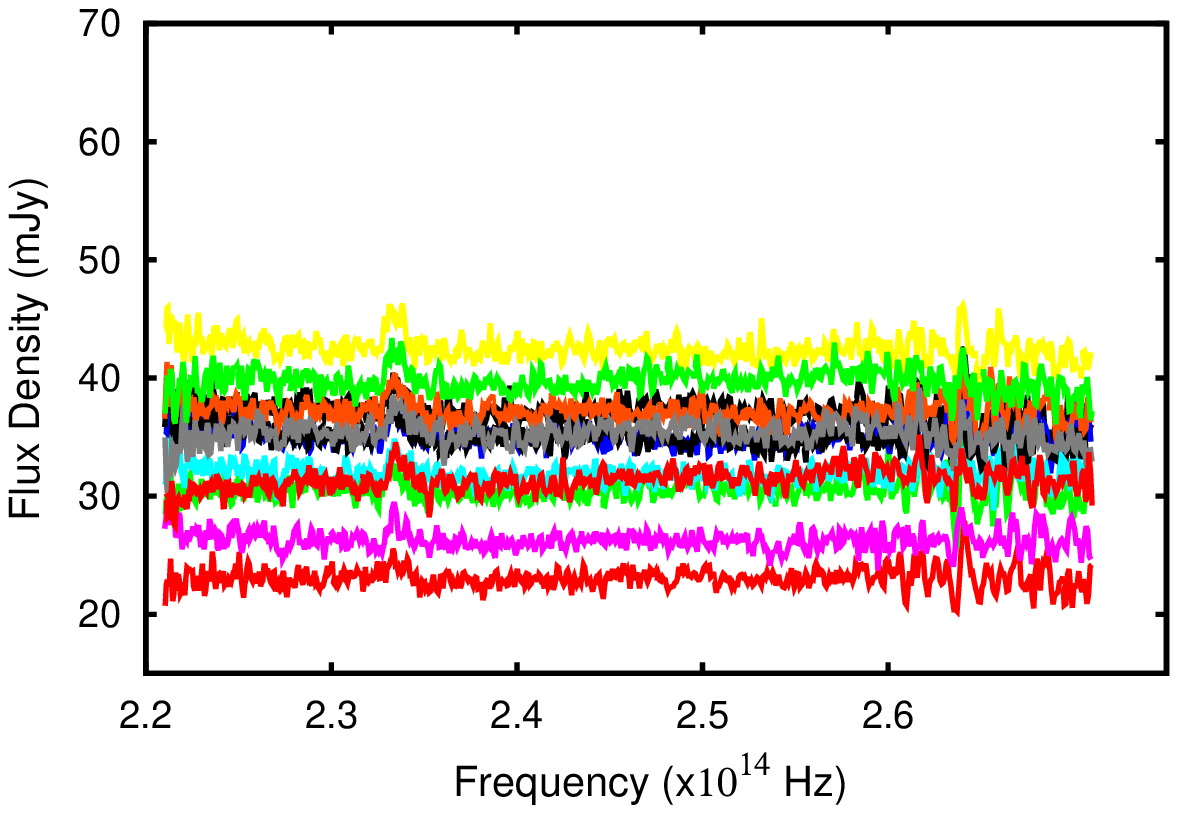}\\
&SS1&\\
\includegraphics[width=5.6cm]{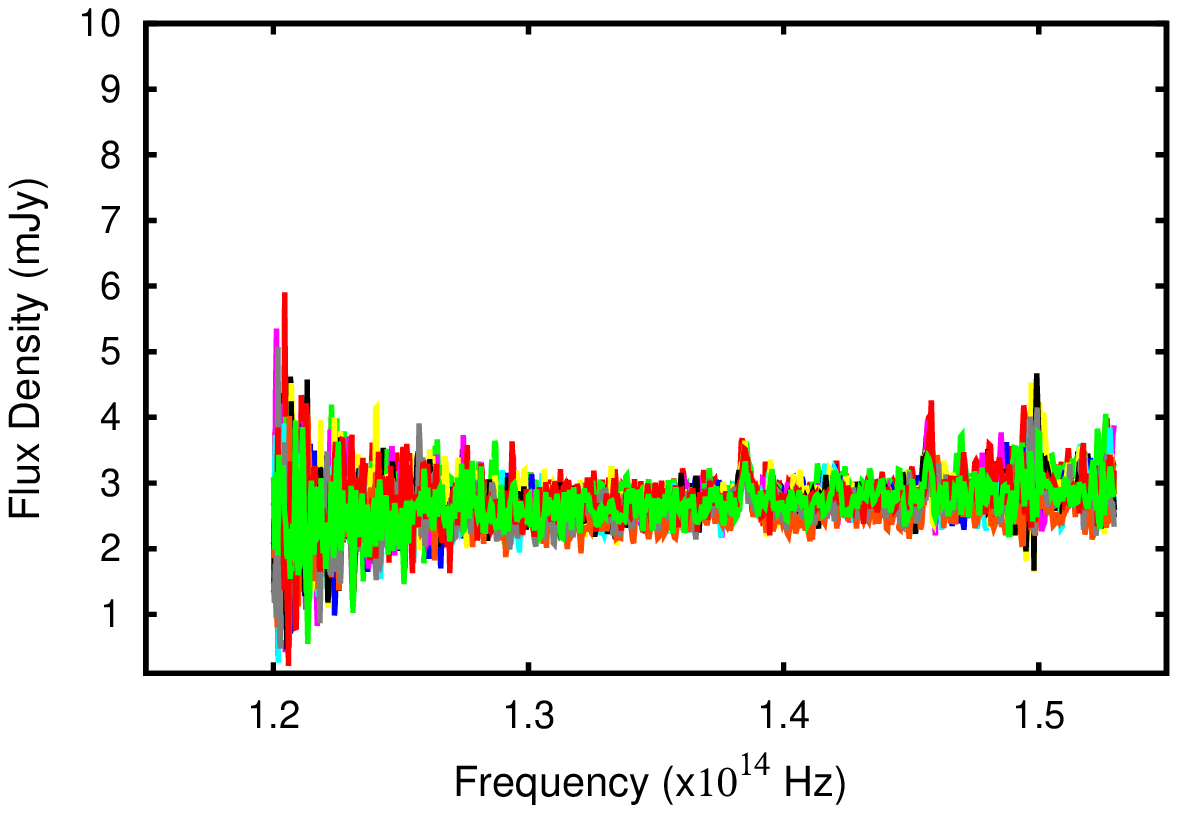}&\includegraphics[width=5.6cm]{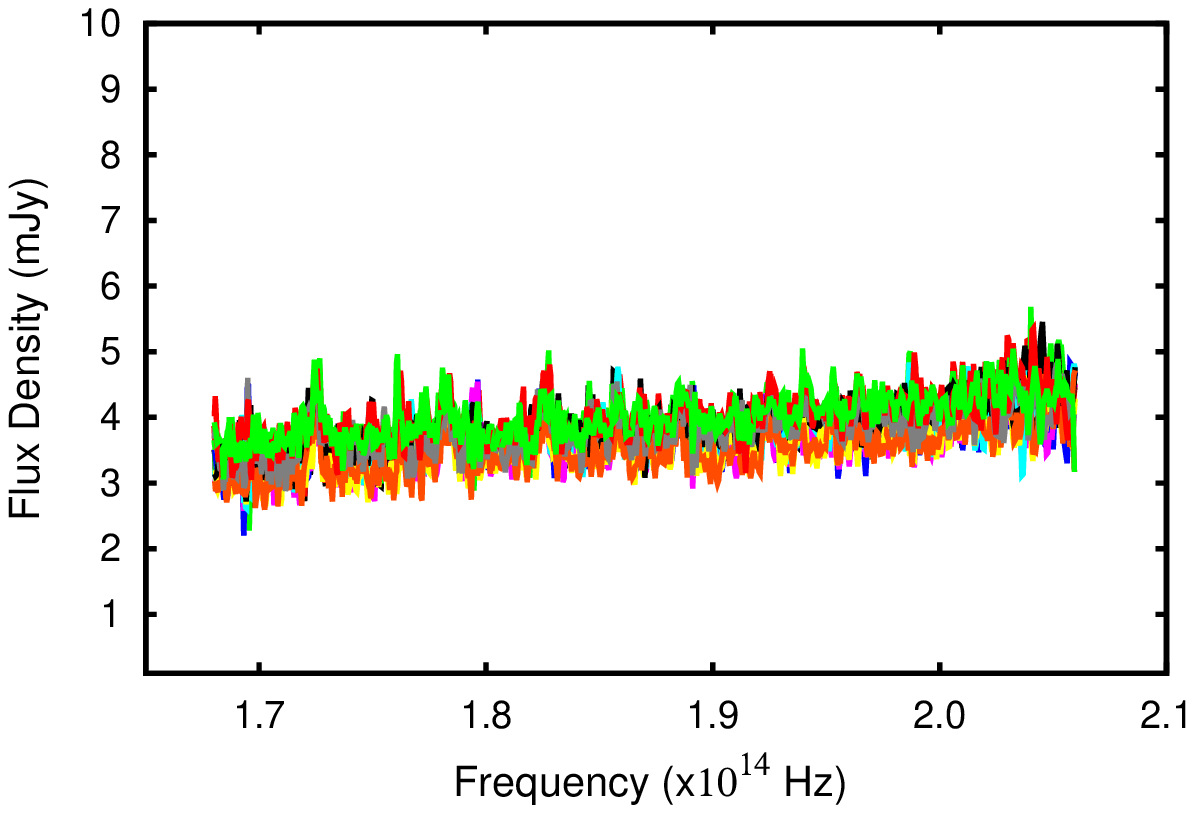}&\includegraphics[width=5.6cm]{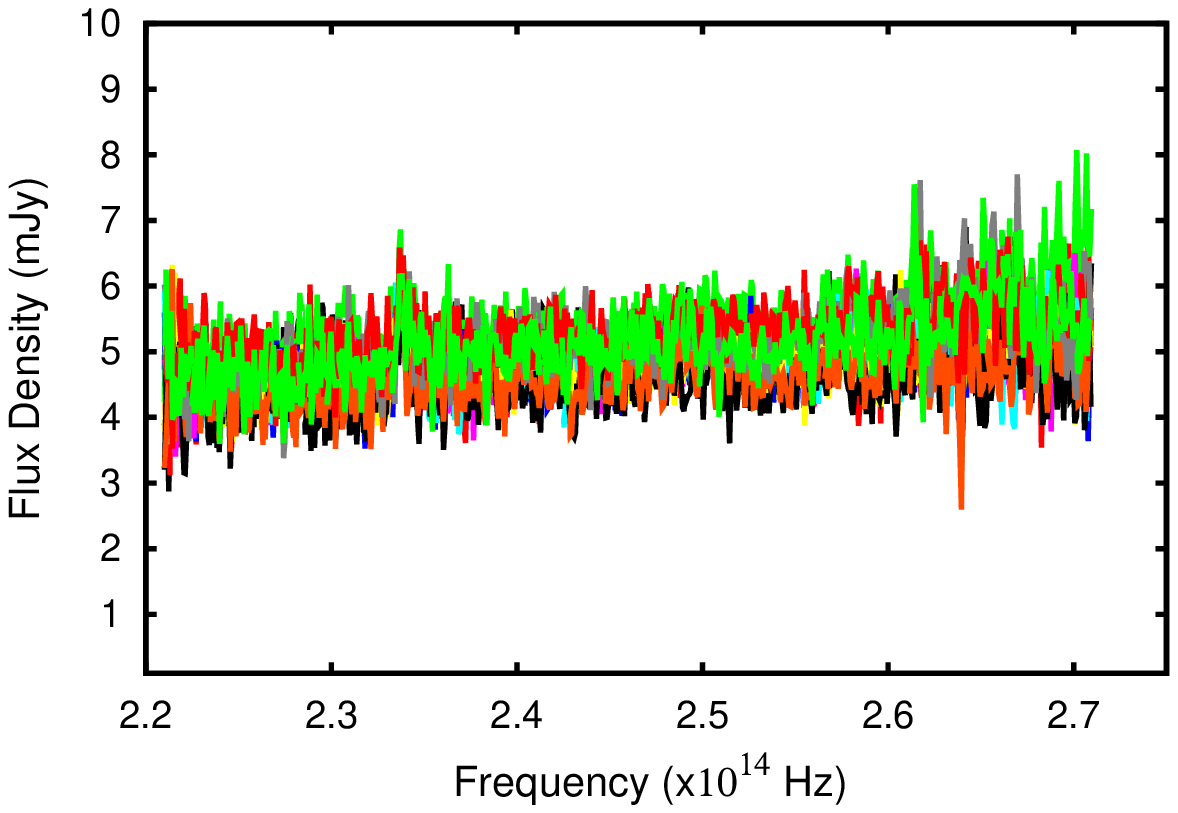}\\
\end{tabular}
\caption{\small Near-IR sub-spectra for HS1 (top), HS2 (middle) and SS1
  (bottom), and in each band ({\it K} to {\it J}, left to right)
  of \gx. Each sub-spectrum corresponds to a 20~s integration.}
\label{timespec}
\end{center}
\end{figure*}
The \rxte\ data were reduced using the \texttt{HEASOFT}~v.~6.11
software package, following the standard procedure described in the \rxte\
cookbook\footnote{http://heasarc.gsfc.nasa.gov/docs/xte/data\_analysis.html}. 
Spectra were extracted from the Proportional
Counter Array \citep[PCA;][]{2006Jahoda}, in the 3--50~keV range. We
only used the top layer of the Proportional Counter Unit (PCU) 2 as it
is the best-calibrated detector out of the five PCUs. Systematic
errors of 0.5 per cent were added to all channels.

We reduced the data from the X-ray
telescope (XRT) on board \sw\ with the \texttt{xrtpipeline}~v.~0.12.6
using standard quality cuts and event grades 0D2. The data
were collected in windowed timing (WT) mode to avoid pile-up. We used
\texttt{xselect}~v.~2.4 to extract source and background spectra using square boxes as
extraction region, in the range 0.5--10~keV. The boxes have a size of 40 pixels along the only
spatial dimension available in WT mode. We generated the ancillary
response file (ARF) with \texttt{xrtmkarf} and used the latest version
(v.~012) of the response matrices provided by the \sw\
team. We finally rebinned the spectrum to get a minimum of 20 counts
per channel.

\subsubsection{Spectral state classification}

To identify the spectral state of \gx\
during Obs.~1, 2, and 3 we fitted, with {\tt XSPEC}~v.~12.7, the
three high-energy spectra with a model combining the contributions
from the accretion disc \citep[\textsc{diskbb,}][]{1984Mitsuda}, an
iron line around 6.4~keV ({\sc Gaussian}), as well as a power law, all modified
by photoelectric extinction \citep[{\sc phabs}, abundances and
cross-sections from][]{1989Anders, 1992Balucinska}. When
needed, a reflection component {\sc 
  reflect} was added for an inclination angle fixed to the most
recent measured value, $46^{\circ}$ \citep{2011Shidatsu}. However, it is
worth remembering that several other measurements between 10 and
50$^{\circ}$ exist \citep[see Table~6 in][for a review]{2011Shidatsu},
but the use of a $10^{\circ}$ inclination angle does not change
significantly the fits. Moreover, the disc's
contribution during Obs.~2 was too low to be detected with
PCA so that {\sc diskbb} was not used in the modelling. The best-fit parameters are listed
in Table~\ref{parobsx}. The results unambiguously show that \gx\ is in
the HS during Obs.~1 and Obs.~2 (hereafter HS1 and HS2) , and in the
SS during Obs.~3 (hereafter SS1). 

\section{Nature and properties of the OIR emission}

We use spectral fitting to understand the origin of the
OIR continuum of \gx. To that purpose, we built the SEDs
for the three observations with all the available data listed
in Table~\ref{logobs}. The OIR spectra were corrected for ISM
extinction along the LOS with the law derived in
\citet{1999Fitzpatrick}. Among the several consistent values for the
optical absorption \citep[see e.g][]{2003Buxton, 2004Hynes}, we adopted
$\Ave\,=\,3.25\pm0.50$ \citep[see][and reference therein]{2011Gandhi},
but it is however worth remembering that it is not well constrained and may be higher
\citep[e.g. $\Ave\,=\,3.7\pm0.3,$][]{1998Zdzi}. In adopting the lower
\Av, our subsequent results, especially those related to the
disc emission in the optical, should be considered as lower
limits. Throughout HS1 and HS2, the source was detected in
the radio domain with a relatively flat emission and we therefore
expect a strong contribution from the compact jets
\citep{2003Corbel}. In contrast, the continuum should be dominated by
the accretion disc during SS1.

\subsection{Near-IR continuum variations at short timescales}

Figure~\ref{timespec} displays, for the three observations and in
each near-IR band, the twelve deredenned ``sub-spectra'' that were 
all obtained with a good signal-to-noise ratio (S/N). The comparison of
their respective continuum may therefore give reliable information
about the intrinsic spectral variations of \gx\ at timescales at least as short
as 20~s, both in the HS and SS. However, the optical spectra being the result
of one exposure only, their time-dependent behaviour cannot be known.

During SS1, the near-IR emission of \gx\ is constant in each band over
the $12\times20$~s exposures. On the contrary, in the HS the emission is
strongly variable in all bands on timescales as low as 20~s. While
this behaviour is consistent 
with the near-IR flickering already observed through photometry
\citep{2010Casella}, it could also be the result of
changing observing conditions, in particular variable slit-losses due
to a variable seeing. However, the seeing was always lower than the
1\arcsec\ slit width, and the flux-calibration we reached during
Obs.~1 is consistent within 5\% with the ESO/REM photometry
\citep{2011Cadolle}; this very likely points towards low
slit-losses. Moreover, the atmospheric conditions were each time
stable over the whole integration. Indeed, the
airmass-corrected seeing at central wavelength, measured by ESO, was
clustered in the ranges 0\farcs62--0\farcs73, 0\farcs63--0\farcs74,
and 0\farcs57--0\farcs66 during Obs.~1, Obs.~2, and Obs.~3,
respectively. When considered per filter, the maximum-to-minimum
variations were always lower than 10\%, and even 5\% 
during Obs.~1. This by itself does not exclude slit-loss variations,
but it is reasonable to assume that their effect is weak. Finally, the
fact that the continuum is constant during SS1 and variable during HS1
and HS2 despite similar atmospheric and instrumental conditions
strengthens the intrinsic origin of the variations in the HS.

That said, the average maximum-to-minimum flux ratios $f_{\rm
    max}/f_{\rm min}$ in {\it J}, {\it H}, and {\it K} are
  $1.36\pm0.08$, $1.42\pm0.08$, and $1.72\pm0.13$ during HS1, as well
  as $1.84\pm0.12$, $1.60\pm0.10$, and $2.29\pm0.17$ during HS2,
  respectively. Because {\it J} precedes {\it H} which itself precedes
  {\it K}, it is not relevant to compare the variations in a filter with
respect to one another; it is however still possible to compare both hard
states. In {\it J} and {\it H}, the continuum is clearly more variable
during HS2. In {\it K}, although HS2 exhibits a higher
maximum-to-minimum flux ratio, it only stems from three sub-spectra
and $f_{\rm max}/f_{\rm  min}=1.15\pm0.04$ otherwise. Overall, the
variations are therefore bluer in HS2 than in HS1, in agreement with their
respective average continuum (see Sect.~3.3 and Figure~\ref{diffobs}); this
likely points towards two different origins.

\subsection{The average OIR continuum during SS1}
\begin{figure}
%\begin{center}
\includegraphics[width=9cm]{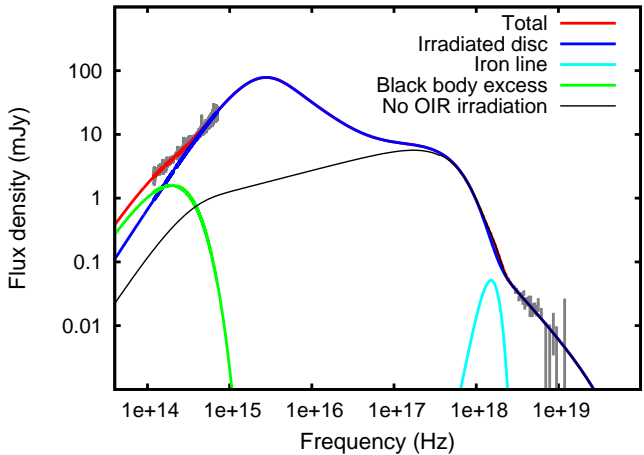}
\caption{\small ISAAC+FORS2+PCA SED of \gx\ during SS1 fitted with
  the irradiated accretion disc {\sc diskir} model. Besides a
  relativistic iron line needed to fit an X-ray excess around
  6~keV, a near-IR excess that we fitted with a $\sim$3348~K
  black body is present at lower frequencies. We also display the
  irradiated disc with no OIR reprocessing.}
\label{fitobs3}
%\end{center}
\end{figure}

The standard multicolour accretion disc alone is unable to reproduce the
OIR spectrum, and a possible explanation is an excess due to X-ray
reprocessing within the disc. In order to get a quantitative
understanding of this phenomenon, we
considered the {\sc diskir} model \citep{2008Gierlinski,
  2009Gierlinski} combined with a Gaussian, to fit
the OIR to X-ray SED. Roughly, it is an extension of the {\sc diskbb} model
that includes comptonisation of the blackbody disc's photons by a
corona of hot electrons \citep[based on {\sc nthcomp},][]{1996Zdzi},
as well as disc irradiation, both from the inner region of the disc
itself and from the corona. The model has nine
parameters: the disc's temperature $kT_{\rm disc}$ and normalisation
\textit{N} (same as {\sc diskbb}), the photon index $\Gamma$ of the hard X-rays power law
and the temperature $kT_{\rm e}$ of the comptonising photons, the ratio between the corona's
and the disc's luminosity $L_{\rm c}/L_{\rm d}$, the fraction of hard
X-ray emission that illuminates the disc $f_{\rm in}$, the irradiated
radius $R_{\rm irr}$ expressed in terms of the disc inner radius, the
fraction of soft X-ray emission which is thermalised in the outer disc
$f_{\rm out}$, and the logarithm of the outer radius of the disc ${\rm
  log10}(R_{out})$ expressed
as a function of the inner radius. The first seven parameters are
completely defined by the high-energy data while the two latter are
characterised by the OIR ones. We fixed several
parameters: (1) the comptonisation temperature
$kT_{\rm e}$ was unconstrained and pegged to the upper boundary
whatever it was. We fixed it to 250~keV, a value high enough to
mimic a power law like hard X-ray spectrum, (2) $f_{\rm in}$ was fixed to 0.3,
well-suited for the SS of microquasars \citep[ 0.1 in the
HS,][]{1997Poutanen, 2005Ibragimov, 2010Gilfanov} (3) after several
unsuccessful attempts, $R_{\rm irr}$ was fixed to 1.1 as it
systematically pegged to the lower value
\citep{2009Gierlinski}. Moreover, this model alone was 
never able to describe the whole SED due to a near-IR excess. We
therefore added a black body for completion, but it is 
important to note that this excess could have other explanations. The
best-fit parameters are listed in Table~\ref{parobs3} 
and the best-fit SED is displayed in Figure~\ref{fitobs3}. It is clear
that accretion disc irradiation dominates the OIR continuum during
SS1. Nevertheless, the derived proportion of reprocessed X-rays within the disc,
$f_{\rm out}=0.05_{-0.02}^{+0.07}$, is very high and at least one order of magnitude
larger than previously found in XTE~J1817$-330$ \citep{2009Gierlinski}
and twice the value for \grs\ \citep{2010Rahoui}. Although we cannot exclude
that this is due to an overestimation of the optical extinction, no
lower value has ever been confirmed yet. \citet{2004Hynes}
derived $\Ave\ge2.64$ from the interstellar Na doublet, but acknowledged that it
was probably too low because the NaD features were saturated. It is therefore very likely that
either another source of thermal emission contributes to the optical spectrum
of \gx\ during SS1, and/or irradiation is enhanced.
  \ctable[
caption = {\small  Best parameters obtained from the fit to the
   \gx\ ISAAC+FORS2+PCA SED during SS1. The best-fit model is
   {\sc phabs$\times$(diskir+gaussian+bbody}). The errorbars are given
   at the 90\% confidence level}, 
label = parobs3,
]{cc}{
\tnote[a]{Same as {\sc diskbb}}
}{\FL
\FL
Parameters&Best-fit value \ML
\nh\ ($\times10^{22}\,{\rm cm}^{-2}$)&0.58 (fixed)\NN\addlinespace[0.05cm]
$kT_{\rm disc}$ (keV)&0.78$_{-0.01}^{+0.01}$\NN\addlinespace[0.05cm]
$\Gamma$&$2.38_{-0.10}^{+0.10}$\NN\addlinespace[0.05cm]
$kT_{\rm e}$ (keV)&$2.50\times10^2$ (fixed)\NN\addlinespace[0.05cm]
$L_{\rm c}/L_{\rm d}$&$0.12_{-0.03}^{+0.03}$\NN\addlinespace[0.05cm]
$f_{\rm in}$&0.30 (fixed)\NN\addlinespace[0.05cm]
$f_{\rm out}$ &$0.05_{-0.02}^{+0.07}$\NN\addlinespace[0.05cm]
${\rm log}10(R_{\rm out})$ ($R_{\rm in}$)& $4.30_{-0.10}^{+0.09}$\NN\addlinespace[0.05cm]
$N$\tmark[a]&$4.35_{-0.33}^{+0.38}\times10^{3}$\NN\addlinespace[0.05cm]
$E_{\rm iron}$ (keV)&5.97 (fixed)\NN\addlinespace[0.05cm]
$\sigma_{\rm iron}$ (keV)&1.33 (fixed)\NN\addlinespace[0.05cm]
$T_{\rm BB}$ (K)&$3.35_{-0.04}^{+0.04}\times10^{3}$\NN\addlinespace[0.05cm]
$R_{\rm BB}/D_{\rm BB}$ ($R_{\odot}$/kpc)&$1.18_{-0.11}^{+0.05}$\NN\addlinespace[0.05cm]
$\chi_{\rm r}^2$ (d.o.f)&0.97 (4442)\LL
}
\begin{figure*}
\begin{center}
\includegraphics[width=18cm]{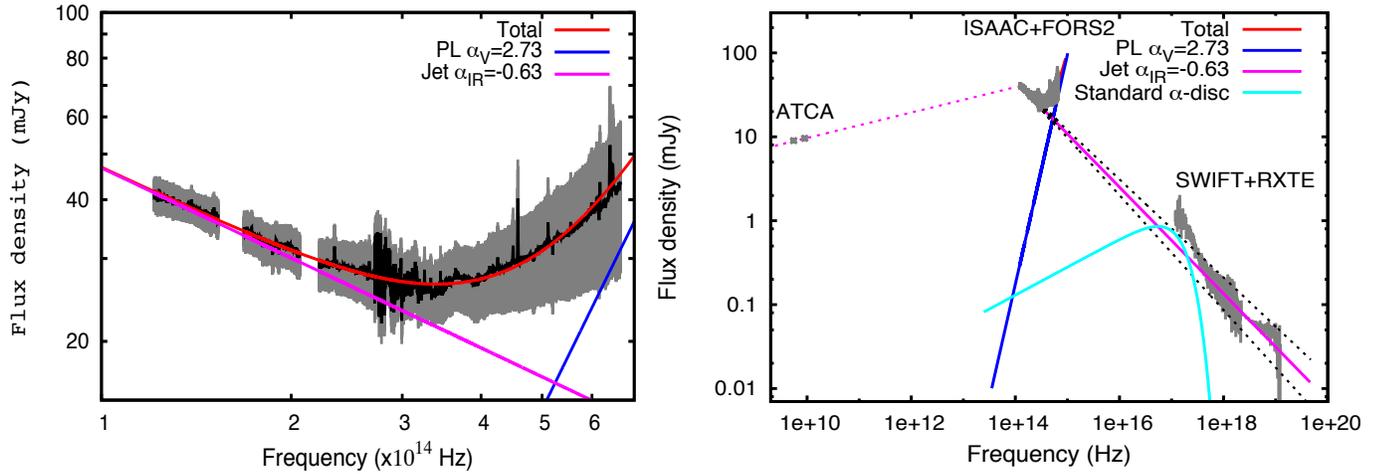}
\caption{\small \textbf{Left}: average OIR continuum during HS1. It is
fitted with a power law mimicking the optically thin synchrotron from
the compact jet ($\alpha_{\rm IR}=-0.63\pm0.05$), combined with a power
law with a spectral index of $\alpha_{\rm V}=2.73\pm0.37$.
\newline
 \textbf{Right}: ATCA+ISAAC+FORS2+XRT+PCA SED of \gx\ during
   HS1. The jet's optically thick power law ($\alpha\approx0.15$,
   magenta dotted-line) is
   extended to the near-IR. The optically thin emission is roughly 
   consistent, within the 3$\sigma$ uncertainties (black dotted-lines) with the
   X-ray spectrum beyond about $6\times10^{17}$~Hz, 
   i.e. 2.5~keV. The accretion disc, derived from the X-ray fitting, is
   also superimposed.}
\label{fitobs1}
\end{center}
\end{figure*}

\subsection{The average OIR continuum during HS1}

\begin{figure*}
\begin{center}
\includegraphics[width=18cm]{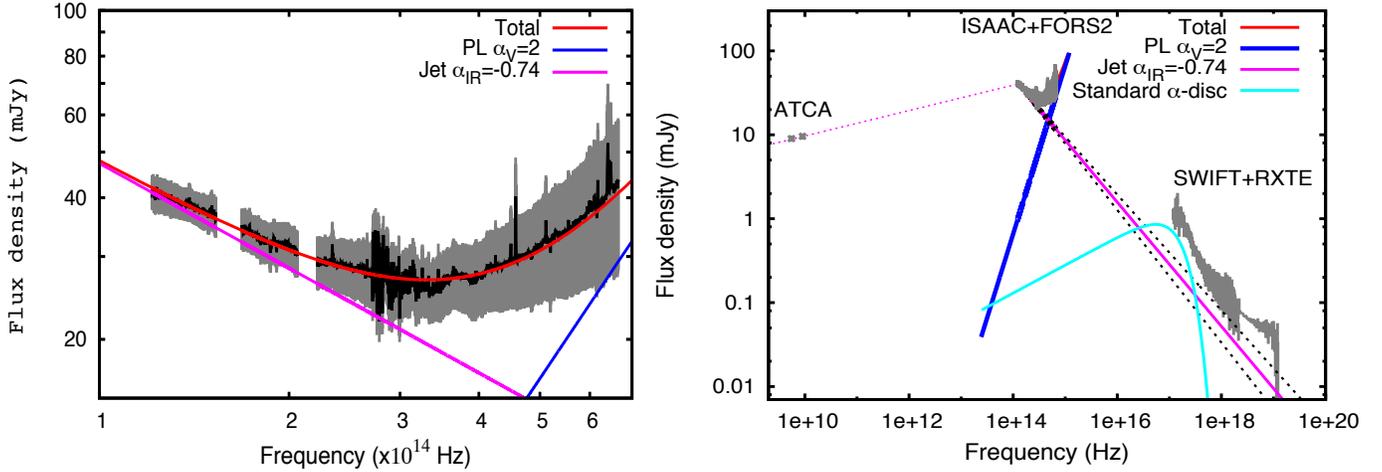}
\caption{\small Same as Figure~\ref{fitobs1} but here, the spectral
  index of the optical power law, $\alpha_{\rm V}$, 
  was fixed to 2 to mimic the Raleigh-Jeans tail of the irradiated
  accretion disc. In this case, the optically thin emission ($\alpha_{\rm IR}=-0.74\pm0.04$) is not
  consistent, within the 3$\sigma$ uncertainties (black dotted-lines)
  with the X-ray spectrum.}
\label{fitobs1alt}
\end{center}
\end{figure*}

Figure~\ref{fitobs1} displays the deredenned OIR spectrum of \gx\ during
HS1 (left panel), as well as the radio to X-ray SED (right
panel). The OIR spectrum is best-described by two power laws whose
spectral indices at a $3\sigma$
  significance are $\alpha_{\rm IR}=-0.63\pm0.05$, typical of
optically thin synchrotron emission from compact jets, and
$\alpha_{\rm V}=2.73\pm0.37$. Moreover the extension, from the
  radio domain, of the optically thick emission ($\alpha\sim0.15$)
  matches the near-IR spectrum, in agreement with a
  spectral break, which we do not detect, located close to
  $1.2\times10^{14}$~Hz \citep[see][]{2011Gandhi}. Likewise, the
extension of the jet optically thin emission is consistent with the 
high energy spectrum beyond about $6\times10^{17}$~Hz, i.e. 2.5~keV
(see right panel, Figure~\ref{fitobs1}). In the optical, the value of $\alpha_{\rm V}$
is larger than the spectral index expected for the Raleigh-Jeans tail of a thermal emission
($\sim2$), which may mean that besides disc irradiation, another
component contributes to the ultraviolet/optical emission of \gx\
during HS1. Now, this could also be due to an
  overestimation of the optical extinction and considering the large optical error bars,
  a good fit is also obtained when $\alpha_{\rm V}$ is fixed
    to 2 (Figure~\ref{fitobs1alt}, left panel). In this case, $\alpha_{\rm IR}=-0.74\pm0.04$, still consistent with optically thin
  synchrotron from the jet, but the extension of the
  jet power law does not match the X-ray spectrum (Figure~\ref{fitobs1alt}, right panel). To perform a more quantitative
    analysis, we used \textsc{diskir} combined with a jet near-IR power law to
describe the OIR/X-ray SED, but the best-fit, although statistically good,
is rather unphysical. Indeed, the fraction of reprocessed X-rays
$f_{\rm out}$ is found to be between 0.1 and 0.5, and the ratio $L_{\rm c}/L_{\rm d}$
between the corona's and disc's luminosities as well as the
temperature of the disc are unconstrained. This either means that the use of
  \textsc{diskir} is not relevant, or that disc irradiation
  combined with the optically thin emission of the jet is not
  sufficient to describe the OIR continuum of \gx\ during HS1.

\subsection{The average OIR continuum during HS2}

\begin{figure}
\begin{center}
\includegraphics[width=9cm]{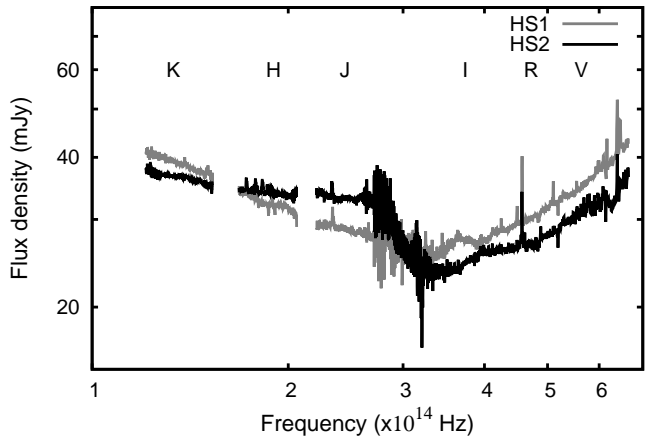}
\caption{\small Comparison between the \gx's OIR
  deredenned spectra, obtained during HS1 (gray) and HS2
  (black).} 
\label{diffobs}
\end{center}
\end{figure}

\begin{figure}
\begin{center}
\includegraphics[width=9cm]{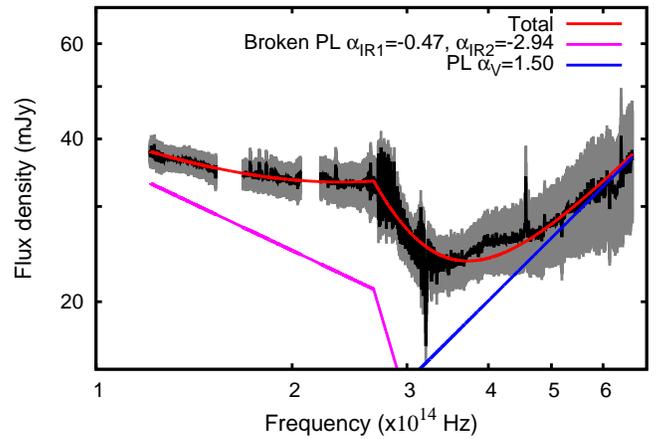}\\
\caption{\small Average OIR continuum during HS2. It is
fitted with a broken power law ($\alpha_{\rm IR1}\approx-0.47$ and $\alpha_{\rm IR2}\approx-2.94$),
combined with a power law with a spectral index fixed to $\alpha_{\rm
  V}=1.5$. Clearly, $\alpha_{\rm IR2}$ is too steep for optically thin
synchrotron radiation.}
\label{fitobs2}
\end{center}
\end{figure}

Figure~\ref{diffobs}, which displays the OIR spectra during HS1
  and HS2,  shows that the two spectra are very different. First, the
  HS2 spectrum is clearly flatter in the near-IR, with a 
continuum lower in {\it K} but higher in {\it H} and {\it J}. Then, 
the flux drops dramatically and it becomes fainter again from {\it
  I}. We can exclude any calibration issue as the explanation since
the level reached in the optical matches very
well to that of the {\it J} band. Moreover, we are not aware of any
FORS2 and/or ISAAC instrumental problem that could have occurred, so that
we are forced to conclude that this drop is real, although the
phenomenon responsible for such peculiar continuum is difficult to
understand. That said, we tested the straightforward explanation,
i.e. the detection of the jet's spectral break, by fitting the spectrum with a model combining 
a broken and a simple power laws. We obtained a good fit, but the
inferred slope ($\alpha\le-2$ beyond the break at
about $2.5\times10^{14}$~Hz) is steeper than expected for optically
thin synchrotron (see Figure~\ref{fitobs2}). This either implies a more complicated emission
mechanism within the jet scenario
and/or another origin for the OIR continuum during HS2.

\section{Discussion}

In the SS, the OIR continuum can be best described by irradiation of
the outer parts of the accretion disc. However, the fit
to the data with an irradiated disc model implies a high 
level of reprocessing, and an excess appears at lower frequencies.
In the HS, not only do the near-IR jet-dominated spectra exhibit rapid
variations, but at least one other process besides disc irradiation and
optically thin synchrotron must be taken into account to explain the spectral continuum. 
\subsection{ \gx\ in the SS} 

\subsubsection{Enhanced irradiation of the accretion disc}

Irradiation of the accretion disc has
long been thought to dominate the UV/optical emission of microquasars
in the SS \citep[see e.g.][]{1990Vrtilek, 1994Paradijsb}, and this is
certainly the case for \gx. 
However, the derived $f_{\rm out}=0.05_{-0.02}^{+0.07}$, although poorly constrained
due to deredenning uncertainties, is at least twice the value
previously measured for \grs\ \citep{2010Rahoui}. It would even be
larger if we considered a higher value for the extinction along the LOS,
\citep[e.g. $\Ave\sim3.7$,][]{1998Zdzi}. We therefore
believe that disc irradiation in \gx\ is enhanced, and two scenarios may be invoked: 
\begin{enumerate}
\item \citet{1996Pringle} and \citet{1996Maloney} showed how
non-axisymmetric radiation forces could induce warping of a flat disc.
In such a configuration, the disc's surface illuminated by the central
source would be larger, and the higher the degree of
warping, the higher the illumination. It is therefore possible that
the high level of X-ray reprocessing observed in
\gx\ is the consequence, at least partially, of a very warped
accretion disc irradiated by the inner regions \citep[see
e.g.][concerning GS~1124$-$68 and A0620$-$00]{2000Esin}.
\item \gx's accretion disc is known to launch winds at
large radii \citep{2004Miller}. Part of the X-ray and/or UV emission
could therefore be up-scattered by these winds back towards the disc
to heat it up locally. This would also enhance the self-irradiation
of the disc, and would be added to the effect of warping \citep[see e.g.][]{1999Dubus}.
\end{enumerate}

\subsubsection{The presence of a near-IR excess}

\gx's emission exhibits a near-IR excess with respect to the
irradiated accretion disc emission. Before proceeding 
further with our interpretation, we acknowledge that it could be
due to discrepancies between the unknown spectral continuum of the
standard star HIP~68124 and the B2/3V synthetic model we made use of
in place to flux-calibrate our data. That said, the excess's magnitudes
in the 2MASS {\it J}, {\it H}, and {\it Ks} filters are $16.93\pm0.34$,
$15.70\pm0.22$, and $14.76\pm0.15$, respectively, approximatively
similar to the lowest published near-IR magnitudes to date 
\citep [$\sim17$, $\sim16$, and $15.2\pm0.3$,][]{2002Chaty, 2003Buxton}. Considering
that the source is in the SS, when the radio activity is quenched
\citep{1999Fenderb}, it is reasonable to assume that a
compact jet contribution is at best marginal. We rather think that
the near-IR excess is due to the secondary and, although a very
simple modelling, we believe that the use of a black body is
appropriate. The companion star has never been identified, but based on the orbital parameters  and
lowest magnitudes in the optical bands, it is thought to be a G, K, or M giant or subgiant
\citep{2001Shahbaz, 2004Hynes, 2004Zdzi, 2008Munoz}. The temperature $T_{\rm
  BB}\sim3348$~K that we derive is consistent with a M star. Moreover, a mass ratio
$q\sim0.125$ \citep{2008Munoz} and a semi-amplitude
$K_{\ast}\sim317$~km~s$^{-1}$ \citep{2004Hynes} lead to
$R_{\ast}/a\sim0.3$ and  $a\sin(i)\sim11.9$~\rsun\ \citep{2004Zdzi},
hence $R_{\ast}\sim3.6/\sin(i)$. The $R_{\rm BB}\sim7.1$~\rsun\ value
\citep[for a minimum distance $D_{\rm BB}\sim6$~kpc,][]{2003Hynes}, not unusual
for an M subgiant, is therefore consistent if the inclination
$i\sim30^{\circ}$, in agreement with previous estimates. This 
implies $a\sim23.8$~\rsun\ and $R_{\rm out}\sim1.2$~\rsun, which 
again are consistent with the orbital parameters of the system. 
However, \citet{2001Shahbaz} measured 
  $r=20.1\pm0.1$ in the off-state, and
  claimed that the undetected companion star was responsible for
  30\% of the flux at maximum, which gives $r\sim21.3$. The inferred
  magnitude of the black body is $r\sim20.4$. While this
  seems to invalidate the companion star hypothesis, such difference
  could actually be explained by orbital variability
  \citep[e.g. 0.4 magnitude in the $I$ band,][]{2002Cowley}, companion
  star irradiation by the accretion disc, and,
  more importantly, by the discrepancies in the optical between a cool
  star spectrum and a black body. We therefore believe that this excess is
indeed due to a M subgiant companion star, but further near-
and mid-IR spectroscopic observations of the source in the off-state
are necessary for confirmation.

\subsubsection{The HS1 case}

There is a strong contribution of the optically thin
synchrotron emission of the jet to the OIR
continuum. While this emission seems to extend to the X-ray domain,
whether or not it means that the jet is responsible for the hard
X-ray emission is discussed below. This result
is not new and was first given in \citet{2002Corbel}. Nevertheless,
the use of spectroscopic OIR data here instead of photometric ones brings better
constraints on the continuum, especially in the optical.  As already
mentioned, accretion disc irradiation as modelled with the
\textsc{diskir} model cannot alone explain the level of
optical emission, as the derived fraction of reprocessed
X-rays is 10--50\%, against 5\% in the SS. Indeed, \citet{1996Jong}
found that a maximum of a few percent of the incoming photons were
absorbed by the accretion disc, and it is hard to believe that disc warping and/or wind
up-scattering could increase the fraction to such a level. Moreover, the
spectral index of the optical power law may be too high compared to that
expected for the Raleigh-Jeans tail of a thermal process, although our
data do not allow us to be definitive. We therefore believe that besides irradiation, there must be
some other physical, perhaps non-thermal, process to account for the
OIR excess in the emission of \gx. A self-consistent model including a
more complex jet and/or accretion physics is beyond the scope of this
paper. Here, we mention two processes that we believe to be the best
candidates, at least in the case of \gx:

\begin{enumerate}
\item the first is pre-shock synchrotron as
described in the framework of the jet model presented in
\citet{2005Markoff}. It corresponds to the emission of a magnetised
reservoir of electrons coming from the accretion flow (similar to a corona), that remain
in a thermal distribution at the base of the jet before they are
accelerated into power laws. This region can produce a significant
UV/OIR emission, whose intensity and peak frequency roughly depend on
the region's size and the distance within the jet from where the
acceleration starts, 
\item the second, recently proposed in \citet{2011Veledina}, also
  involves synchrotron radiation from electrons located in a corona close to the
  BH. Nonetheless, the population has a non-thermal distribution. This
  region produces a strong OIR emission, whose intensity and
  peak frequency mainly depend on its size,  and an X-ray/OIR anti-correlation
  is found. Combined with the time-lagging X-ray/OIR
  correlation due to X-ray reprocessing within the disc,
  this model can explain the pattern observed in the optical/X-ray
  cross-correlation function of \gx\ \citep{2008Gandhi}. 
\end{enumerate} 

\subsubsection{The HS2 case}
\subsection{\gx\ in the HS} 

We are unsure of the phenomenon responsible for such
peculiar continuum. A possible interpretation is that it is
due to the same process that might be responsible, 
in addition to disc irradiation, for the optical excess during
HS1. Indeed, we may be witnessing a shift of this component to
the lower frequencies, where it peaks in {\it H} or {\it J}. This would explain the flatter
slope of the optical excess, more consistent with disc
irradiation. Such shift is predicted for the two processes described
in Sect.~4.2.1. For instance, \citet{2009Maitra}, using the jet model presented in
\citet{2005Markoff}, showed how changes in the emitting region's size
and/or the distance from which acceleration starts led to pre-shock
synchrotron peaking either in the UV/optical or the near-IR (Figure~4
in their paper). However, another explanation could be a more complex compact
jet's spectral continuum, such as those presented in
\citet{2009Peer}. In their paper, the authors take adiabatic losses
into account, both for power law and Maxwellian
distributions. Depending on the value of the magnetic field, they show
that this gives a large sample of continuum shapes. In particular,
Figures~6, 7, and 8 in their paper, which displays profiles corresponding to a
Maxwellian distribution for different magnetic field values, could
correspond to the compact jet emission during HS2.  

\subsubsection{Understanding the rapid near-IR continuum variability}

During HS1, the variability must be mainly driven by rapid fluctuations in the
compact jet, and it is interesting that this phenomenon has almost no
effect on the slope. This is a hint that the optically thin synchrotron emission
is rapidly moving upwards or downwards. This may imply rapid changes
in the location -- in flux and frequency -- 
of the undetected spectral break \citep{2011Gandhi}, i.e. significant variations 
at timescales at least as low as 20~s of the physical conditions at
the base of the jet, such as the magnetic field and/or the base
radius. Moreover, the fact that during HS1, the
optical, {\it  J}, {\it H}, and {\it K} spectra, although
non-simultaneous, are all consistent with each other and that the
extension of the near-IR jet power law to the 
high energy domain roughly describes the 3--50~keV \rxte\ spectrum is
surprising. It means that the optical and near-IR
sub-second flickering observed in the HS \citep[se e.g.][]{2010Gandhi,
  2010Casella} are smoothed out and that we observe the average
behaviour of the compact jet beyond about a few hundred seconds. This
is in good agreement with the existence of tight near-IR/X-ray 
correlations in microquasars in general and \gx\ 
in particular \citep[see e.g.][]{2004Gleissner, 2009Coriat}. In any case,
in a recent study of \cygx1\ \citep{2011Rahoui}, we argue that 
the extension of the optically thin synchrotron emission from the
spectral break we detect in the mid-IR to the $\gamma-$ray domain
matches pretty well to the ten-year averaged polarised
emission mentioned in \citet{2011Laurent}. In contrast, the contribution of the
compact jets to the X-ray spectrum (\rxte/PCA+HEXTE) is found
marginal, a hint that it probably comes from a hot corona. In this
context, a confirmation of the domination of compact jets in the \gx's
X-ray emission could mean that both microquasars belong to
two distinct families, perhaps differentiated by the
mass-accretion rate level and/or luminosity \citep{2010Russell,
  2011Coriat, 2011Sobo}. 

Considering that we do not clearly understand the origin of the near-IR
emission of \gx\ during HS2, it is difficult to speculate about
the observed variations. The continuum is more variable in {\it H} and {\it J} than during
HS1 and,  although {\it K} exhibits a higher maximum-to-minimum
flux ratio, it only stems from three sub-spectra and $f_{\rm max}/f_{\rm
  min} =1.15\pm0.04$ otherwise. This behaviour could mean that the
variations are not driven by the same processes, or that the compact
jet contributes more to the {\it H} and {\it J} bands, in other terms
that the spectral break is located in the near-IR. It is also puzzling that the
discrepancies observed in the OIR between both HS have no
equivalent in the \rxte/PCA spectra. The reason could be that the
X-ray data were obtained more than half a day after the OIR ones, i.e. an
event could have occurred while \gx\ was being observed in the OIR
before returning to the ``normal'' during the
\rxte\ observations. But it could also mean that the hard X-ray
emission is due to another component, and whatever happens to the jets has no
consequence in the high energy domain, at least not at these timescales. 
\section{Conclusion}

We presented a comprehensive multi-wavelength study of \gx\ in
the SS and HS, with a focus on the OIR spectral continuum.  We showed 
that the SS is dominated by enhanced irradiation of the accretion
disc, with an additional contribution from a M subgiant companion star. In
one of the HS, the OIR emission stems from the compact jet, disc
irradiation, and maybe synchrotron from the inner regions. The second
HS spectrum exhibits a peculiar continuum which could be a hint of a
more complex emission. In both cases, the near-IR spectrum is strongly
variable on timescales at least as short as 20~s, which we suggested is a
manifestation of rapid variations of the magnetic field and/or nozzle radius. Our
results highlight the complexity of the OIR emission of a
microquasar like \gx, and a good understanding of the source's
behaviour in this spectral domain requires further in-depth spectroscopic
observations as well as theoretical modelling. In particular, the use
of an instrument allowing simultaneous integrations in the whole OIR range, such as
ESO/X-shooter, would be of great help to investigate the variations -- at long and short timescales -- of the
continuum in a more robust way. Finally, this strengthens the need of the James Webb Space
Telescope ({\it JWST}) for microquasar studies, as the unmatched
sensitivity of NIRSpec and MIRI will allow a simultaneous spectroscopic coverage from 1 to
30~\mic. 

\section*{Acknowledgments}
We thank the anonymous referee for his/her useful comments. FR thanks the ESO staff who performed the service observations. FR and JCL
thanks the Harvard Faculty of Arts and Sciences and the Harvard College
Observatory. MCB acknowledges support from the
Faculty of the European Space Astronomy Centre (ESAC). JAT
acknowledges partial support from NASA under {\em Swift} 
Guest Observer grant NNX10AF94G. SC and JR acknowledges
partial funding from the European Community’s Seventh Framework
Programme (FP7/2007-2013) under grant agreement number ITN 215212
``Black Hole Universe''. DMR acknowledges support from the Netherlands
Organisation for Scientific Research (NWO) Veni Fellowship. This
research has made use of data obtained from the High Energy
Astrophysics Science Archive Research Center (HEASARC), provided by
NASA's Goddard Space Flight Center. The Australia Telescope Compact
Array is part of the Australia Telescope which is funded by the
Commonwealth of Australia for operation as a National Facility managed by
CSIRO. This research has made use of NASA's Astrophysics Data
System, of the SIMBAD, and VizieR databases operated at CDS, 
Strasbourg, France.
\bibliographystyle{mn2e}

\bibliography{./mybib}{}

\begin{thebibliography}{}

\bibitem[\protect\citeauthoryear{{Anders} \& {Grevesse}}{{Anders} \&
  {Grevesse}}{1989}]{1989Anders}
{Anders} E.,  {Grevesse} N.,  1989, GeCoA, 53, 197

\bibitem[\protect\citeauthoryear{{Balucinska-Church} \&
  {McCammon}}{{Balucinska-Church} \& {McCammon}}{1992}]{1992Balucinska}
{Balucinska-Church} M.,  {McCammon} D.,  1992, ApJ, 400, 699

\bibitem[\protect\citeauthoryear{{Blandford} \& {Konigl}}{{Blandford} \&
  {Konigl}}{1979}]{1979Blandford}
{Blandford} R.~D.,  {Konigl} A.,  1979, ApJ, 232, 34

\bibitem[\protect\citeauthoryear{{Buxton} \& {Vennes}}{{Buxton} \&
  {Vennes}}{2003}]{2003Buxton}
{Buxton} M.,  {Vennes} S.,  2003, MNRAS, 342, 105

\bibitem[\protect\citeauthoryear{{Buxton} \& {Bailyn}}{{Buxton} \&
  {Bailyn}}{2004}]{2004Buxton}
{Buxton} M.~M.,  {Bailyn} C.~D.,  2004, The Astronomer's Telegram, 316, 1

\bibitem[\protect\citeauthoryear{{Cadolle Bel}, {Rodriguez}, {D'Avanzo},
  {Russell}, {Tomsick}, {Corbel}, {Lewis}, {Rahoui}, {Buxton}, {Goldoni} \&
  {Kuulkers}}{{Cadolle Bel} et~al.}{2011}]{2011Cadolle}
{Cadolle Bel} M.,  {Rodriguez} J.,  {D'Avanzo} P.,  {Russell} D.~M.,  {Tomsick}
  J.,  {Corbel} S.,  {Lewis} F.~W.,  {Rahoui} F.,  {Buxton} M.,  {Goldoni} P.,
    {Kuulkers} E.,  2011, A\&A, 534, A119+

\bibitem[\protect\citeauthoryear{{Casella}, {Maccarone}, {O'Brien}, {Fender},
  {Russell}, {van der Klis}, {Pe'Er}, {Maitra}, {Altamirano}, {Belloni},
  {Kanbach}, {Klein-Wolt}, {Mason}, {Soleri}, {Stefanescu}, {Wiersema} \&
  {Wijnands}}{{Casella} et~al.}{2010}]{2010Casella}
{Casella} P.,  {Maccarone} T.~J.,  {O'Brien} K.,  {Fender} R.~P.,  {Russell}
  D.~M.,  {van der Klis} M.,  {Pe'Er} A.,  {Maitra} D.,  {Altamirano} D.,
  {Belloni} T.,  {Kanbach} G.,  {Klein-Wolt} M.,  {Mason} E.,  {Soleri} P.,
  {Stefanescu} A.,  {Wiersema} K.,    {Wijnands} R.,  2010, MNRAS, 404, L21

\bibitem[\protect\citeauthoryear{{Chaty}, {Mirabel}, {Goldoni}, {Mereghetti},
  {Duc}, {Mart{\'{\i}}} \& {Mignani}}{{Chaty} et~al.}{2002}]{2002Chaty}
{Chaty} S.,  {Mirabel} I.~F.,  {Goldoni} P.,  {Mereghetti} S.,  {Duc} P.-A.,
  {Mart{\'{\i}}} J.,    {Mignani} R.~P.,  2002, MNRAS, 331, 1065

\bibitem[\protect\citeauthoryear{{Corbel} \& {Fender}}{{Corbel} \&
  {Fender}}{2002}]{2002Corbel}
{Corbel} S.,  {Fender} R.~P.,  2002, ApJL, 573, L35

\bibitem[\protect\citeauthoryear{{Corbel}, {Nowak}, {Fender}, {Tzioumis} \&
  {Markoff}}{{Corbel} et~al.}{2003}]{2003Corbel}
{Corbel} S.,  {Nowak} M.~A.,  {Fender} R.~P.,  {Tzioumis} A.~K.,    {Markoff}
  S.,  2003, A\&A, 400, 1007

\bibitem[\protect\citeauthoryear{{Coriat}, {Corbel}, {Buxton}, {Bailyn},
  {Tomsick}, {K{\"o}rding} \& {Kalemci}}{{Coriat} et~al.}{2009}]{2009Coriat}
{Coriat} M.,  {Corbel} S.,  {Buxton} M.~M.,  {Bailyn} C.~D.,  {Tomsick} J.~A.,
  {K{\"o}rding} E.,    {Kalemci} E.,  2009, MNRAS, 400, 123

\bibitem[\protect\citeauthoryear{{Coriat}, {Corbel}, {Prat}, {Miller-Jones},
  {Cseh}, {Tzioumis}, {Brocksopp}, {Rodriguez}, {Fender} \&
  {Sivakoff}}{{Coriat} et~al.}{2011}]{2011Coriat}
{Coriat} M.,  {Corbel} S.,  {Prat} L.,  {Miller-Jones} J.~C.~A.,  {Cseh} D.,
  {Tzioumis} A.~K.,  {Brocksopp} C.,  {Rodriguez} J.,  {Fender} R.~P.,
  {Sivakoff} G.~R.,  2011, MNRAS, 414, 677

\bibitem[\protect\citeauthoryear{{Cowley}, {Schmidtke}, {Hutchings} \&
  {Crampton}}{{Cowley} et~al.}{2002}]{2002Cowley}
{Cowley} A.~P.,  {Schmidtke} P.~C.,  {Hutchings} J.~B.,    {Crampton} D.,
  2002, AJ, 123, 1741

\bibitem[\protect\citeauthoryear{{Cramer}}{{Cramer}}{1984}]{1984Cramer}
{Cramer} N.,  1984, A\&A, 132, 283

\bibitem[\protect\citeauthoryear{{de Jong}, {van Paradijs} \& {Augusteijn}}{{de
  Jong} et~al.}{1996}]{1996Jong}
{de Jong} J.~A.,  {van Paradijs} J.,    {Augusteijn} T.,  1996, A\&A, 314, 484

\bibitem[\protect\citeauthoryear{{Dubus}, {Lasota}, {Hameury} \&
  {Charles}}{{Dubus} et~al.}{1999}]{1999Dubus}
{Dubus} G.,  {Lasota} J.-P.,  {Hameury} J.-M.,    {Charles} P.,  1999, MNRAS,
  303, 139

\bibitem[\protect\citeauthoryear{{Esin}, {Kuulkers}, {McClintock} \&
  {Narayan}}{{Esin} et~al.}{2000}]{2000Esin}
{Esin} A.~A.,  {Kuulkers} E.,  {McClintock} J.~E.,    {Narayan} R.,  2000, ApJ,
  532, 1069

\bibitem[\protect\citeauthoryear{{Falcke} \& {Biermann}}{{Falcke} \&
  {Biermann}}{1995}]{1995Falcke}
{Falcke} H.,  {Biermann} P.~L.,  1995, A\&A, 293, 665

\bibitem[\protect\citeauthoryear{{Fender}, {Corbel}, {Tzioumis}, {McIntyre},
  {Campbell-Wilson}, {Nowak}, {Sood}, {Hunstead}, {Harmon}, {Durouchoux} \&
  {Heindl}}{{Fender} et~al.}{1999}]{1999Fenderb}
{Fender} R.,  {Corbel} S.,  {Tzioumis} T.,  {McIntyre} V.,  {Campbell-Wilson}
  D.,  {Nowak} M.,  {Sood} R.,  {Hunstead} R.,  {Harmon} A.,  {Durouchoux} P.,
    {Heindl} W.,  1999, ApJL, 519, L165

\bibitem[\protect\citeauthoryear{{Fitzpatrick}}{{Fitzpatrick}}{1999}]{1999Fitzpatrick}
{Fitzpatrick} E.~L.,  1999, PASP, 111, 63

\bibitem[\protect\citeauthoryear{{Gandhi}, {Blain}, {Russell}, {Casella},
  {Malzac}, {Corbel}, {D'Avanzo}, {Lewis}, {Markoff}, {Cadolle Bel}, {Goldoni},
  {Wachter}, {Khangulyan} \& {Mainzer}}{{Gandhi} et~al.}{2011}]{2011Gandhi}
{Gandhi} P.,  {Blain} A.~W.,  {Russell} D.~M.,  {Casella} P.,  {Malzac} J.,
  {Corbel} S.,  {D'Avanzo} P.,  {Lewis} F.~W.,  {Markoff} S.,  {Cadolle Bel}
  M.,  {Goldoni} P.,  {Wachter} S.,  {Khangulyan} D.,    {Mainzer} A.,  2011,
  ApJL, 740, L13+

\bibitem[\protect\citeauthoryear{{Gandhi}, {Dhillon}, {Durant}, {Fabian},
  {Kubota}, {Makishima}, {Malzac}, {Marsh}, {Miller}, {Shahbaz}, {Spruit} \&
  {Casella}}{{Gandhi} et~al.}{2010}]{2010Gandhi}
{Gandhi} P.,  {Dhillon} V.~S.,  {Durant} M.,  {Fabian} A.~C.,  {Kubota} A.,
  {Makishima} K.,  {Malzac} J.,  {Marsh} T.~R.,  {Miller} J.~M.,  {Shahbaz} T.,
   {Spruit} H.~C.,    {Casella} P.,  2010, MNRAS, 407, 2166

\bibitem[\protect\citeauthoryear{{Gandhi}, {Makishima}, {Durant}, {Fabian},
  {Dhillon}, {Marsh}, {Miller}, {Shahbaz} \& {Spruit}}{{Gandhi}
  et~al.}{2008}]{2008Gandhi}
{Gandhi} P.,  {Makishima} K.,  {Durant} M.,  {Fabian} A.~C.,  {Dhillon} V.~S.,
  {Marsh} T.~R.,  {Miller} J.~M.,  {Shahbaz} T.,    {Spruit} H.~C.,  2008,
  MNRAS, 390, L29

\bibitem[\protect\citeauthoryear{{Gierli{\'n}ski}, {Done} \&
  {Page}}{{Gierli{\'n}ski} et~al.}{2008}]{2008Gierlinski}
{Gierli{\'n}ski} M.,  {Done} C.,    {Page} K.,  2008, MNRAS, 388, 753

\bibitem[\protect\citeauthoryear{{Gierli{\'n}ski}, {Done} \&
  {Page}}{{Gierli{\'n}ski} et~al.}{2009}]{2009Gierlinski}
{Gierli{\'n}ski} M.,  {Done} C.,    {Page} K.,  2009, MNRAS, 392, 1106

\bibitem[\protect\citeauthoryear{{Gilfanov}}{{Gilfanov}}{2010}]{2010Gilfanov}
{Gilfanov} M.,  2010, in {T.~Belloni} ed., Lecture Notes in Physics, Berlin
  Springer Verlag Vol.~794 of Lecture Notes in Physics, Berlin Springer Verlag,
  {X-Ray Emission from Black-Hole Binaries}.
pp 17--+

\bibitem[\protect\citeauthoryear{{Gleissner}, {Wilms}, {Pooley}, {Nowak},
  {Pottschmidt}, {Markoff}, {Heinz}, {Klein-Wolt}, {Fender} \&
  {Staubert}}{{Gleissner} et~al.}{2004}]{2004Gleissner}
{Gleissner} T.,  {Wilms} J.,  {Pooley} G.~G.,  {Nowak} M.~A.,  {Pottschmidt}
  K.,  {Markoff} S.,  {Heinz} S.,  {Klein-Wolt} M.,  {Fender} R.~P.,
  {Staubert} R.,  2004, A\&A, 425, 1061

\bibitem[\protect\citeauthoryear{{Homan}, {Buxton}, {Markoff}, {Bailyn},
  {Nespoli} \& {Belloni}}{{Homan} et~al.}{2005}]{2005Homanb}
{Homan} J.,  {Buxton} M.,  {Markoff} S.,  {Bailyn} C.~D.,  {Nespoli} E.,
  {Belloni} T.,  2005, ApJ, 624, 295

\bibitem[\protect\citeauthoryear{{Hynes}, {Steeghs}, {Casares}, {Charles} \&
  {O'Brien}}{{Hynes} et~al.}{2003}]{2003Hynes}
{Hynes} R.~I.,  {Steeghs} D.,  {Casares} J.,  {Charles} P.~A.,    {O'Brien} K.,
   2003, ApJL, 583, L95

\bibitem[\protect\citeauthoryear{{Hynes}, {Steeghs}, {Casares}, {Charles} \&
  {O'Brien}}{{Hynes} et~al.}{2004}]{2004Hynes}
{Hynes} R.~I.,  {Steeghs} D.,  {Casares} J.,  {Charles} P.~A.,    {O'Brien} K.,
   2004, ApJ, 609, 317

\bibitem[\protect\citeauthoryear{{Ibragimov}, {Poutanen}, {Gilfanov},
  {Zdziarski} \& {Shrader}}{{Ibragimov} et~al.}{2005}]{2005Ibragimov}
{Ibragimov} A.,  {Poutanen} J.,  {Gilfanov} M.,  {Zdziarski} A.~A.,
  {Shrader} C.~R.,  2005, MNRAS, 362, 1435

\bibitem[\protect\citeauthoryear{Jahoda, Markwardt, Radeva, Rots, Stark, Swank,
  Strohmayer \& Zhang}{Jahoda et~al.}{2006}]{2006Jahoda}
Jahoda K.,  Markwardt C.~B.,  Radeva Y.,  Rots A.~H.,  Stark M.~J.,  Swank
  J.~H.,  Strohmayer T.~E.,    Zhang W.,  2006, ApJSS, 163, 401

\bibitem[\protect\citeauthoryear{{Laurent}, {Rodriguez}, {Wilms}, {Cadolle
  Bel}, {Pottschmidt} \& {Grinberg}}{{Laurent} et~al.}{2011}]{2011Laurent}
{Laurent} P.,  {Rodriguez} J.,  {Wilms} J.,  {Cadolle Bel} M.,  {Pottschmidt}
  K.,    {Grinberg} V.,  2011, Science, 332, 438

\bibitem[\protect\citeauthoryear{{Maitra}, {Markoff}, {Brocksopp}, {Noble},
  {Nowak} \& {Wilms}}{{Maitra} et~al.}{2009}]{2009Maitra}
{Maitra} D.,  {Markoff} S.,  {Brocksopp} C.,  {Noble} M.,  {Nowak} M.,
  {Wilms} J.,  2009, MNRAS, 398, 1638

\bibitem[\protect\citeauthoryear{{Maloney}, {Begelman} \& {Pringle}}{{Maloney}
  et~al.}{1996}]{1996Maloney}
{Maloney} P.~R.,  {Begelman} M.~C.,    {Pringle} J.~E.,  1996, ApJ, 472, 582

\bibitem[\protect\citeauthoryear{{Markoff}, {Nowak} \& {Wilms}}{{Markoff}
  et~al.}{2005}]{2005Markoff}
{Markoff} S.,  {Nowak} M.~A.,    {Wilms} J.,  2005, ApJ, 635, 1203

\bibitem[\protect\citeauthoryear{{Migliari}, {Tomsick}, {Maccarone}, {Gallo},
  {Fender}, {Nelemans} \& {Russell}}{{Migliari} et~al.}{2006}]{2006Migliari}
{Migliari} S.,  {Tomsick} J.~A.,  {Maccarone} T.~J.,  {Gallo} E.,  {Fender}
  R.~P.,  {Nelemans} G.,    {Russell} D.~M.,  2006, ApJL, 643, L41

\bibitem[\protect\citeauthoryear{{Migliari}, {Tomsick}, {Miller-Jones},
  {Heinz}, {Hynes}, {Fender}, {Gallo}, {Jonker} \& {Maccarone}}{{Migliari}
  et~al.}{2010}]{2010Migliari}
{Migliari} S.,  {Tomsick} J.~A.,  {Miller-Jones} J.~C.~A.,  {Heinz} S.,
  {Hynes} R.~I.,  {Fender} R.~P.,  {Gallo} E.,  {Jonker} P.~G.,    {Maccarone}
  T.~J.,  2010, ApJ, 710, 117

\bibitem[\protect\citeauthoryear{{Miller}, {Raymond}, {Fabian}, {Homan},
  {Nowak}, {Wijnands}, {van der Klis}, {Belloni}, {Tomsick}, {Smith}, {Charles}
  \& {Lewin}}{{Miller} et~al.}{2004}]{2004Miller}
{Miller} J.~M.,  {Raymond} J.,  {Fabian} A.~C.,  {Homan} J.,  {Nowak} M.~A.,
  {Wijnands} R.,  {van der Klis} M.,  {Belloni} T.,  {Tomsick} J.~A.,  {Smith}
  D.~M.,  {Charles} P.~A.,    {Lewin} W.~H.~G.,  2004, ApJ, 601, 450

\bibitem[\protect\citeauthoryear{{Mitsuda}, {Inoue}, {Koyama}, {Makishima},
  {Matsuoka}, {Ogawara}, {Suzuki}, {Tanaka}, {Shibazaki} \& {Hirano}}{{Mitsuda}
  et~al.}{1984}]{1984Mitsuda}
{Mitsuda} K.,  {Inoue} H.,  {Koyama} K.,  {Makishima} K.,  {Matsuoka} M.,
  {Ogawara} Y.,  {Suzuki} K.,  {Tanaka} Y.,  {Shibazaki} N.,    {Hirano} T.,
  1984, PASJ, 36, 741

\bibitem[\protect\citeauthoryear{{Mu{\~n}oz-Darias}, {Casares} \&
  {Mart{\'{\i}}nez-Pais}}{{Mu{\~n}oz-Darias} et~al.}{2008}]{2008Munoz}
{Mu{\~n}oz-Darias} T.,  {Casares} J.,    {Mart{\'{\i}}nez-Pais} I.~G.,  2008,
  MNRAS, 385, 2205

\bibitem[\protect\citeauthoryear{{Pe'er} \& {Casella}}{{Pe'er} \&
  {Casella}}{2009}]{2009Peer}
{Pe'er} A.,  {Casella} P.,  2009, ApJ, 699, 1919

\bibitem[\protect\citeauthoryear{{Poutanen}, {Krolik} \& {Ryde}}{{Poutanen}
  et~al.}{1997}]{1997Poutanen}
{Poutanen} J.,  {Krolik} J.~H.,    {Ryde} F.,  1997, MNRAS, 292, L21

\bibitem[\protect\citeauthoryear{{Pringle}}{{Pringle}}{1996}]{1996Pringle}
{Pringle} J.~E.,  1996, MNRAS, 281, 357

\bibitem[\protect\citeauthoryear{{Rahoui}, {Chaty}, {Rodriguez}, {Fuchs},
  {Mirabel} \& {Pooley}}{{Rahoui} et~al.}{2010}]{2010Rahoui}
{Rahoui} F.,  {Chaty} S.,  {Rodriguez} J.,  {Fuchs} Y.,  {Mirabel} I.~F.,
  {Pooley} G.~G.,  2010, ApJ, 715, 1191

\bibitem[\protect\citeauthoryear{{Rahoui}, {Lee}, {Heinz}, {Hines},
  {Pottschmidt}, {Wilms} \& {Grinberg}}{{Rahoui} et~al.}{2011}]{2011Rahoui}
{Rahoui} F.,  {Lee} J.~C.,  {Heinz} S.,  {Hines} D.~C.,  {Pottschmidt} K.,
  {Wilms} J.,    {Grinberg} V.,  2011, ApJ, 736, 63

\bibitem[\protect\citeauthoryear{{Russell}, {Curran}, {Mu{\~n}oz-Darias},
  {Lewis}, {Motta}, {Stiele}, {Belloni}, {Miller-Jones}, {Jonker}, {O'Brien},
  {Homan}, {Casella}, {Gandhi}, {Soleri}, {Markoff}, {Maitra} \& {et
  al.}}{{Russell} et~al.}{2011}]{2011Russell}
{Russell} D.~M.,  {Curran} P.~A.,  {Mu{\~n}oz-Darias} T.,  {Lewis} F.,  {Motta}
  S.,  {Stiele} H.,  {Belloni} T.,  {Miller-Jones} J.~C.~A.,  {Jonker} P.~G.,
  {O'Brien} K.,  {Homan} J.,  {Casella} P.,  {Gandhi} P.,  {Soleri} P.,
  {Markoff} S.,  {Maitra} D.,    {et al.} 2011, MNRAS, in press, arXiv
  1109.3654

\bibitem[\protect\citeauthoryear{{Russell}, {Maitra}, {Dunn} \&
  {Markoff}}{{Russell} et~al.}{2010}]{2010Russell}
{Russell} D.~M.,  {Maitra} D.,  {Dunn} R.~J.~H.,    {Markoff} S.,  2010, MNRAS,
  405, 1759

\bibitem[\protect\citeauthoryear{{Shahbaz}, {Fender} \& {Charles}}{{Shahbaz}
  et~al.}{2001}]{2001Shahbaz}
{Shahbaz} T.,  {Fender} R.,    {Charles} P.~A.,  2001, A\&A, 376, L17

\bibitem[\protect\citeauthoryear{{Shidatsu}, {Ueda}, {Tazaki}, {Yoshikawa},
  {Nagayama}, {Nagata}, {Oi}, {Yamaoka}, {Takahashi}, {Kubota}, {Cottam},
  {Remillard} \& {Negoro}}{{Shidatsu} et~al.}{2011}]{2011Shidatsu}
{Shidatsu} M.,  {Ueda} Y.,  {Tazaki} F.,  {Yoshikawa} T.,  {Nagayama} T.,
  {Nagata} T.,  {Oi} N.,  {Yamaoka} K.,  {Takahashi} H.,  {Kubota} A.,
  {Cottam} J.,  {Remillard} R.,    {Negoro} H.,  2011, PASJ, in press, arXiv
  1105.3586

\bibitem[\protect\citeauthoryear{{Sobolewska}, {Papadakis}, {Done} \&
  {Malzac}}{{Sobolewska} et~al.}{2011}]{2011Sobo}
{Sobolewska} M.~A.,  {Papadakis} I.~E.,  {Done} C.,    {Malzac} J.,  2011,
  MNRAS, 417, 280

\bibitem[\protect\citeauthoryear{{van Paradijs} \& {McClintock}}{{van Paradijs}
  \& {McClintock}}{1994}]{1994Paradijsb}
{van Paradijs} J.,  {McClintock} J.~E.,  1994, A\&A, 290, 133

\bibitem[\protect\citeauthoryear{{Veledina}, {Poutanen} \& {Vurm}}{{Veledina}
  et~al.}{2011}]{2011Veledina}
{Veledina} A.,  {Poutanen} J.,    {Vurm} I.,  2011, ApJL, 737, L17+

\bibitem[\protect\citeauthoryear{{Vrtilek}, {Raymond}, {Garcia}, {Verbunt},
  {Hasinger} \& {Kurster}}{{Vrtilek} et~al.}{1990}]{1990Vrtilek}
{Vrtilek} S.~D.,  {Raymond} J.~C.,  {Garcia} M.~R.,  {Verbunt} F.,  {Hasinger}
  G.,    {Kurster} M.,  1990, A\&A, 235, 162

\bibitem[\protect\citeauthoryear{{Zdziarski}, {Gierli{\'n}ski},
  {Miko{\l}ajewska}, {Wardzi{\'n}ski}, {Smith}, {Harmon} \&
  {Kitamoto}}{{Zdziarski} et~al.}{2004}]{2004Zdzi}
{Zdziarski} A.~A.,  {Gierli{\'n}ski} M.,  {Miko{\l}ajewska} J.,
  {Wardzi{\'n}ski} G.,  {Smith} D.~M.,  {Harmon} B.~A.,    {Kitamoto} S.,
  2004, MNRAS, 351, 791

\bibitem[\protect\citeauthoryear{{Zdziarski}, {Johnson} \&
  {Magdziarz}}{{Zdziarski} et~al.}{1996}]{1996Zdzi}
{Zdziarski} A.~A.,  {Johnson} W.~N.,    {Magdziarz} P.,  1996, MNRAS, 283, 193

\bibitem[\protect\citeauthoryear{{Zdziarski}, {Poutanen}, {Mikolajewska},
  {Gierlinski}, {Ebisawa} \& {Johnson}}{{Zdziarski} et~al.}{1998}]{1998Zdzi}
{Zdziarski} A.~A.,  {Poutanen} J.,  {Mikolajewska} J.,  {Gierlinski} M.,
  {Ebisawa} K.,    {Johnson} W.~N.,  1998, MNRAS, 301, 435

\end{thebibliography}

\clearpage

\appendix

\section{Flux-calibrated synthetic telluric spectra}
\begin{figure}
\begin{tabular}{l}
\includegraphics[width=9cm]{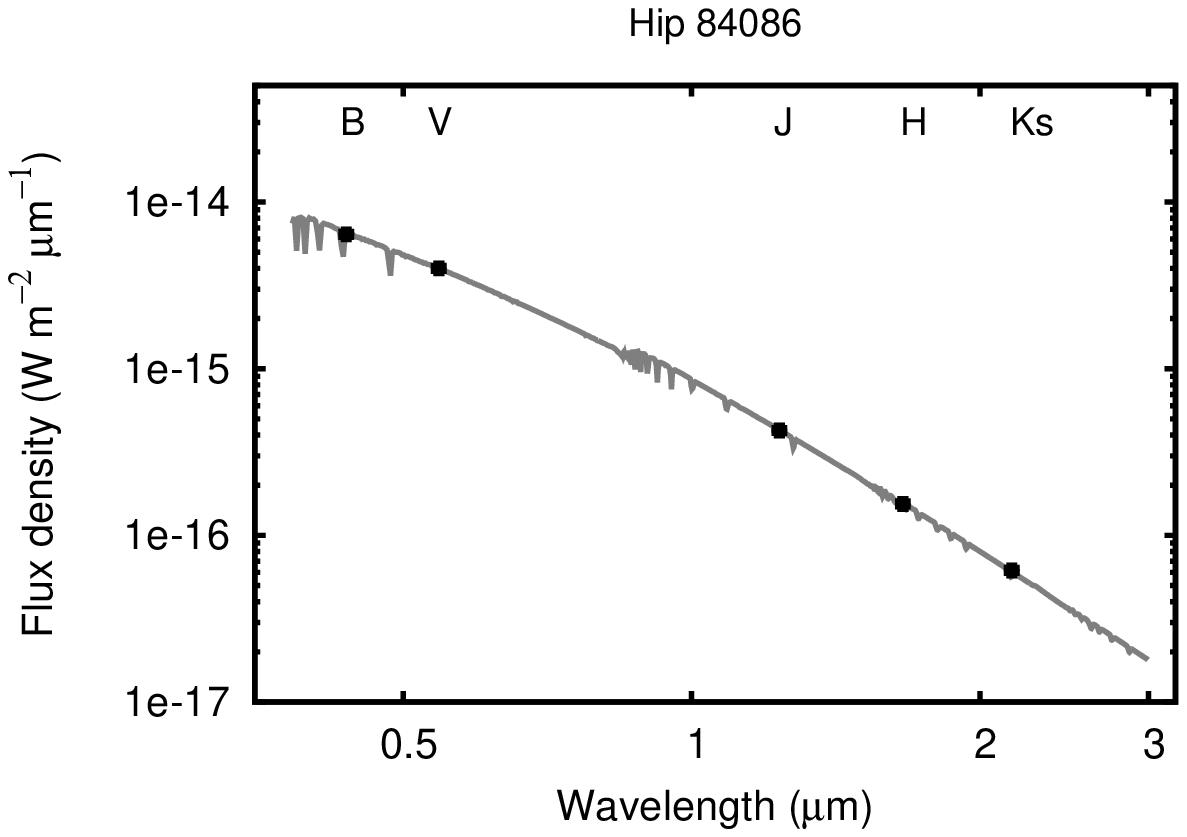}\\
\includegraphics[width=9cm]{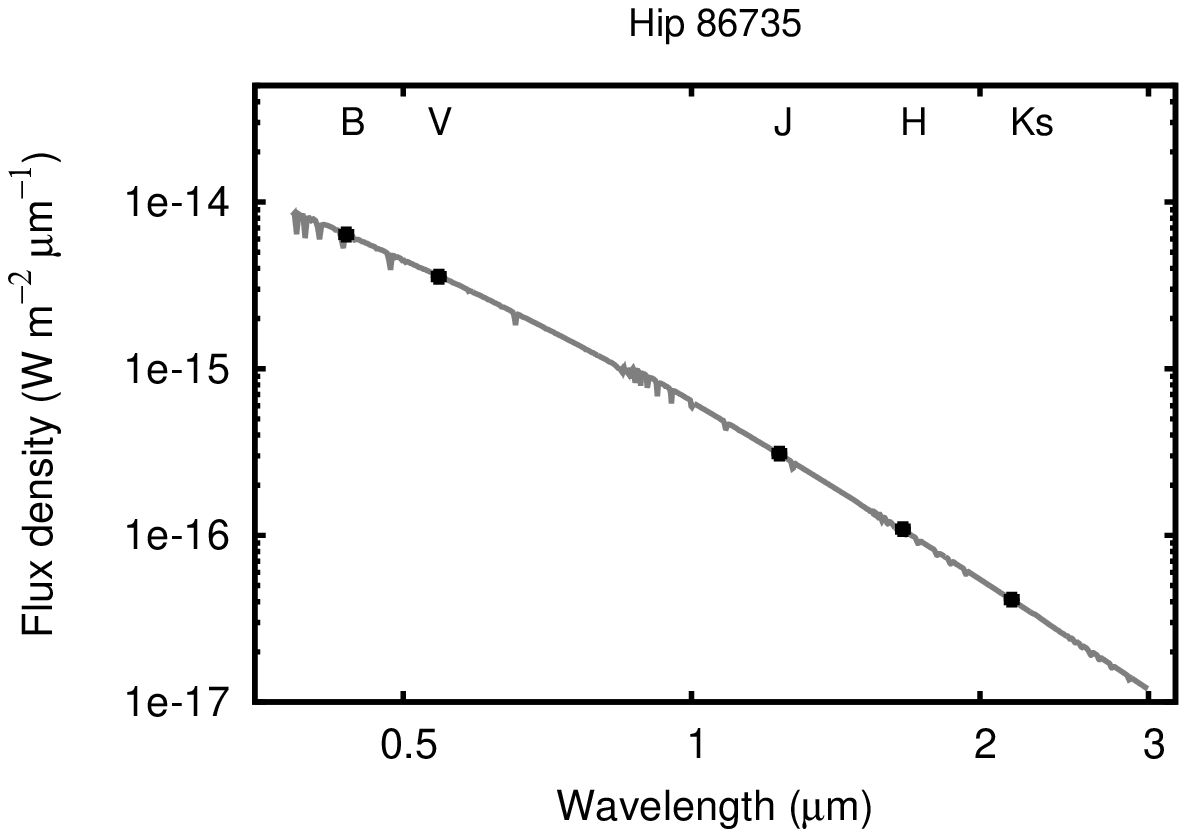}\\
\includegraphics[width=9cm]{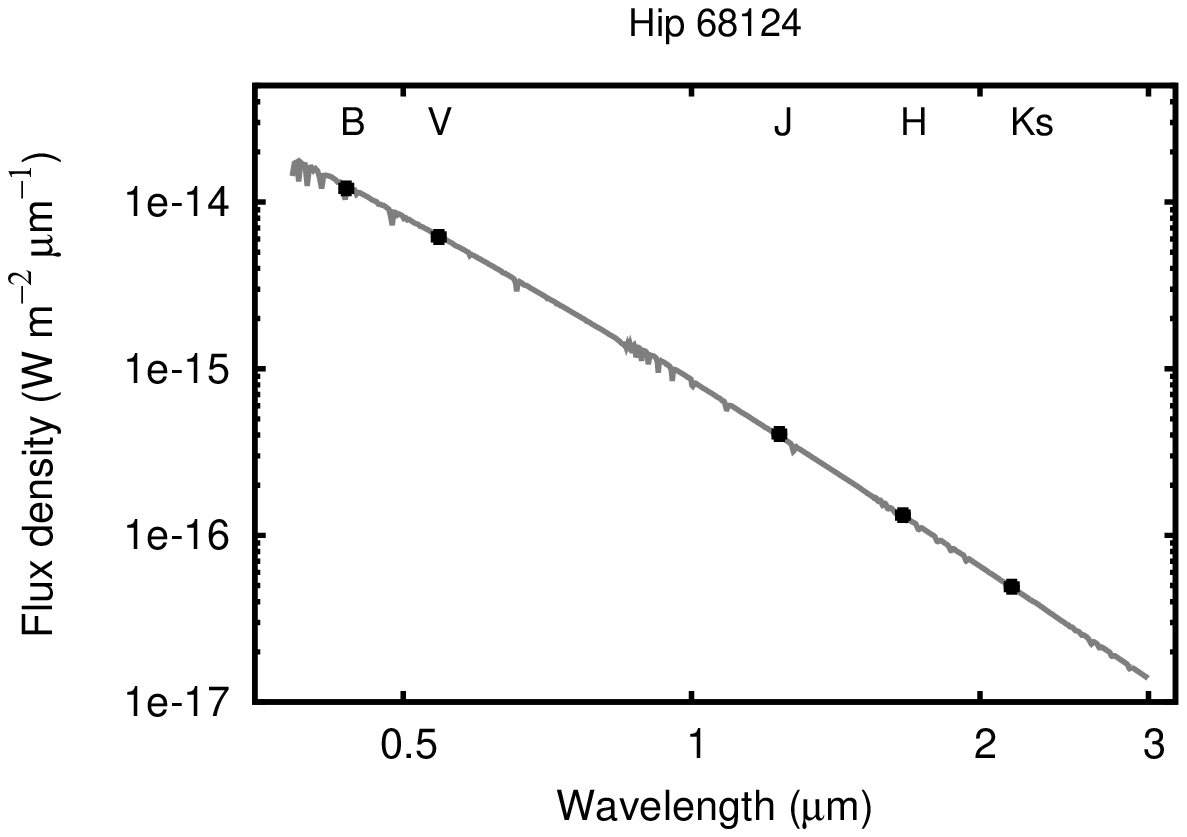}\\
\end{tabular}
\caption{\small Synthetic spectra used for the near-IR spectroscopic
  standard stars in this study. Their measured {\it B}, {\it V}, {\it J}, {\it
    H}, {\it Ks} flux densities are superimposed.}
\label{stdtheo}
\end{figure}

\begin{figure}
\begin{center}
\begin{tabular}{l}
\includegraphics[width=9cm]{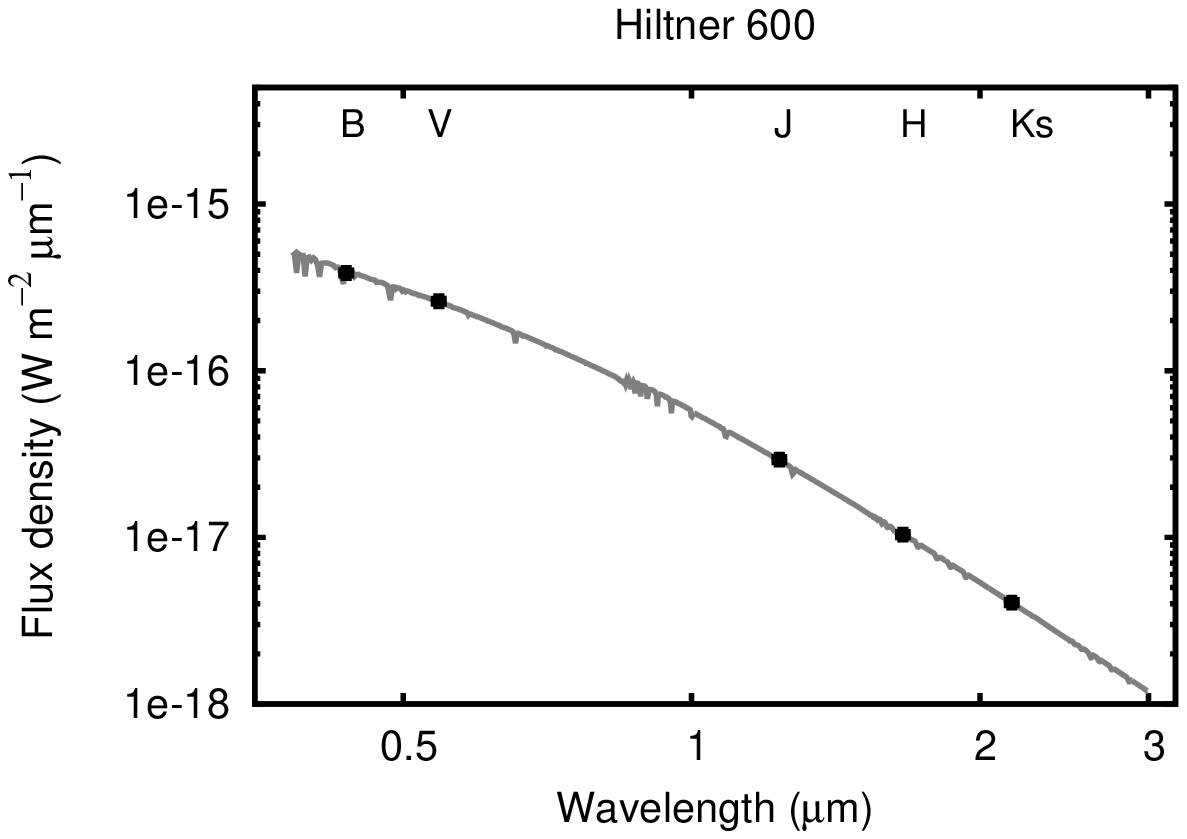}\\
\includegraphics[width=9cm]{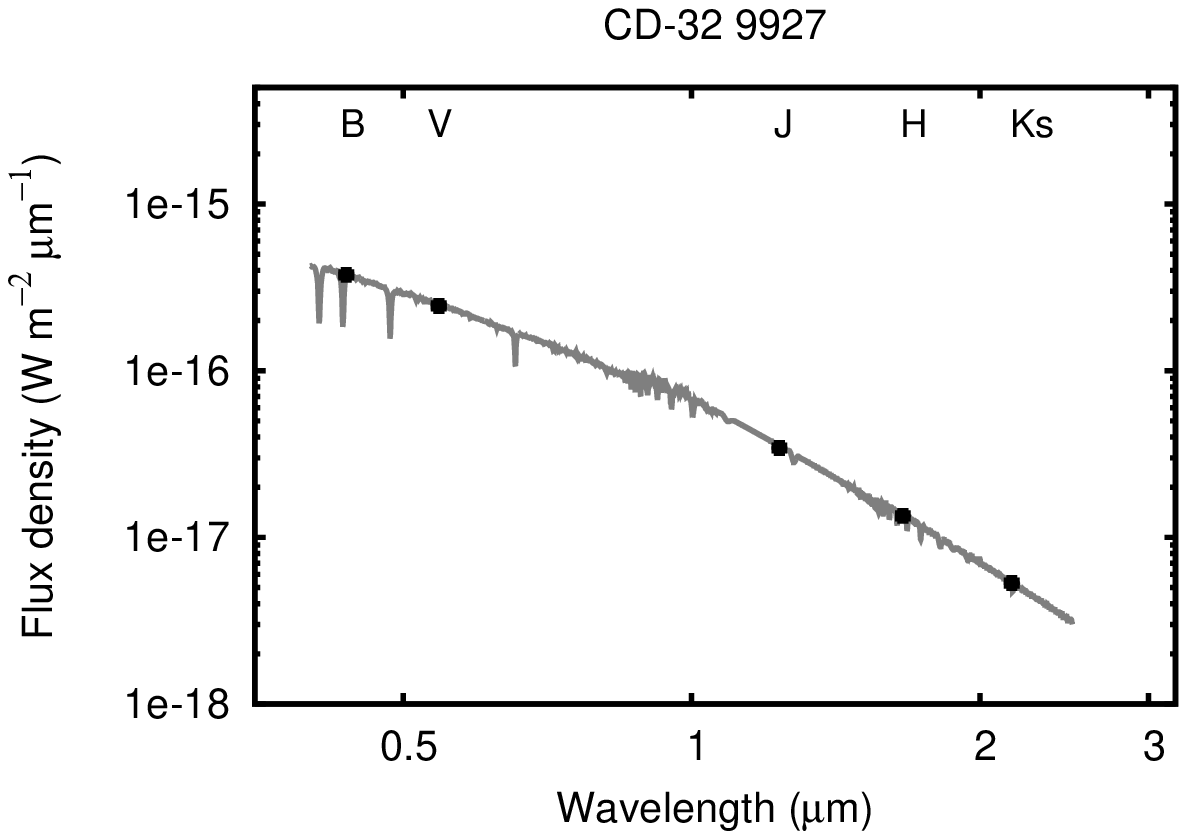}\\
\end{tabular}
\caption{\small Synthetic spectra used for the optical spectroscopic
  standard stars in this study. Their measured {\it B}, {\it V}, {\it J}, {\it
    H}, {\it Ks} flux densities are superimposed.}
\label{stdtheo2}
\end{center}
\end{figure}

In the near-IR, the lack of spectro-photometric standard stars implies the use
of synthetic spectra for flux-calibration purpose. Figure~\ref{stdtheo} displays the
Kurucz spectra we used in the near-IR, and Figure~\ref{stdtheo2} displays the
ones we used in the optical in order to extend the
wavelength coverage beyond 1.02~\mic.

\section{Comparison of \gx's optical spectra derived from
the two methods}

\begin{figure*}
\begin{center}
\begin{tabular}{cc}
\includegraphics[width=9cm]{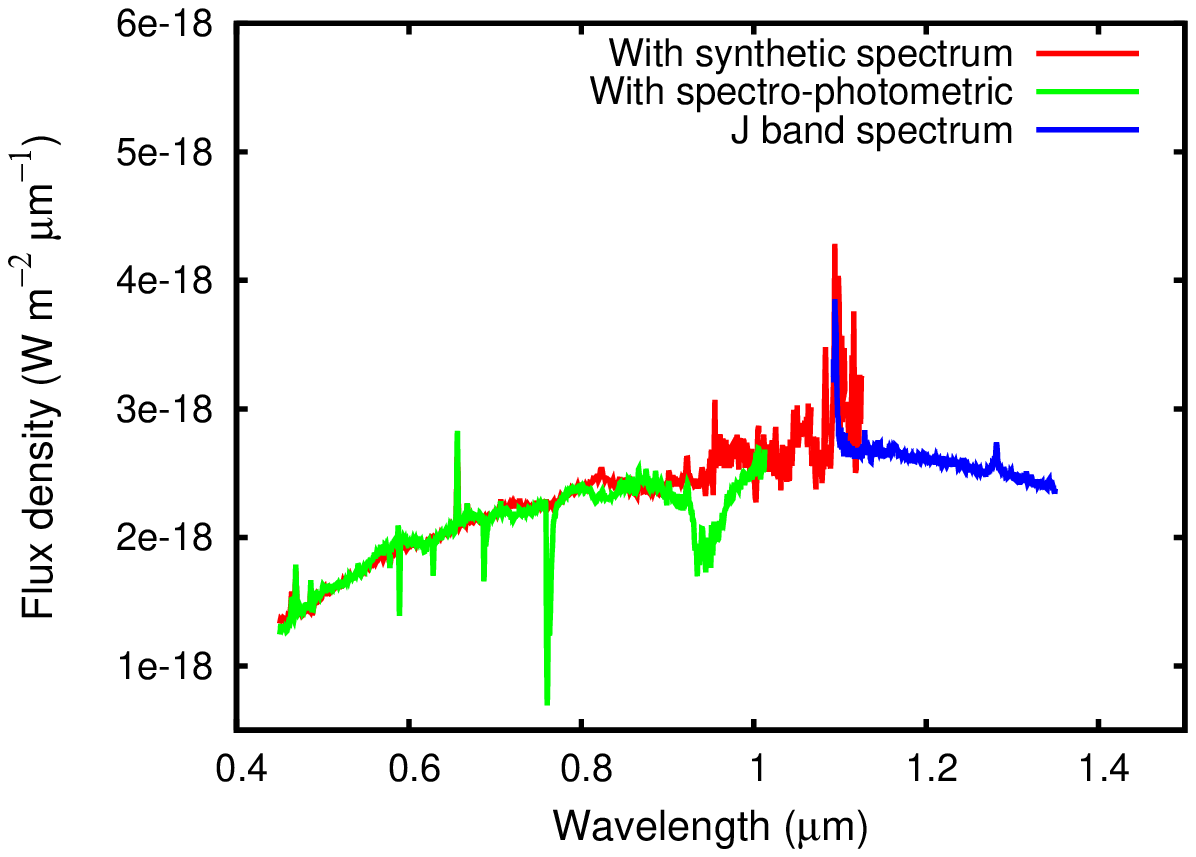}&\includegraphics[width=9cm]{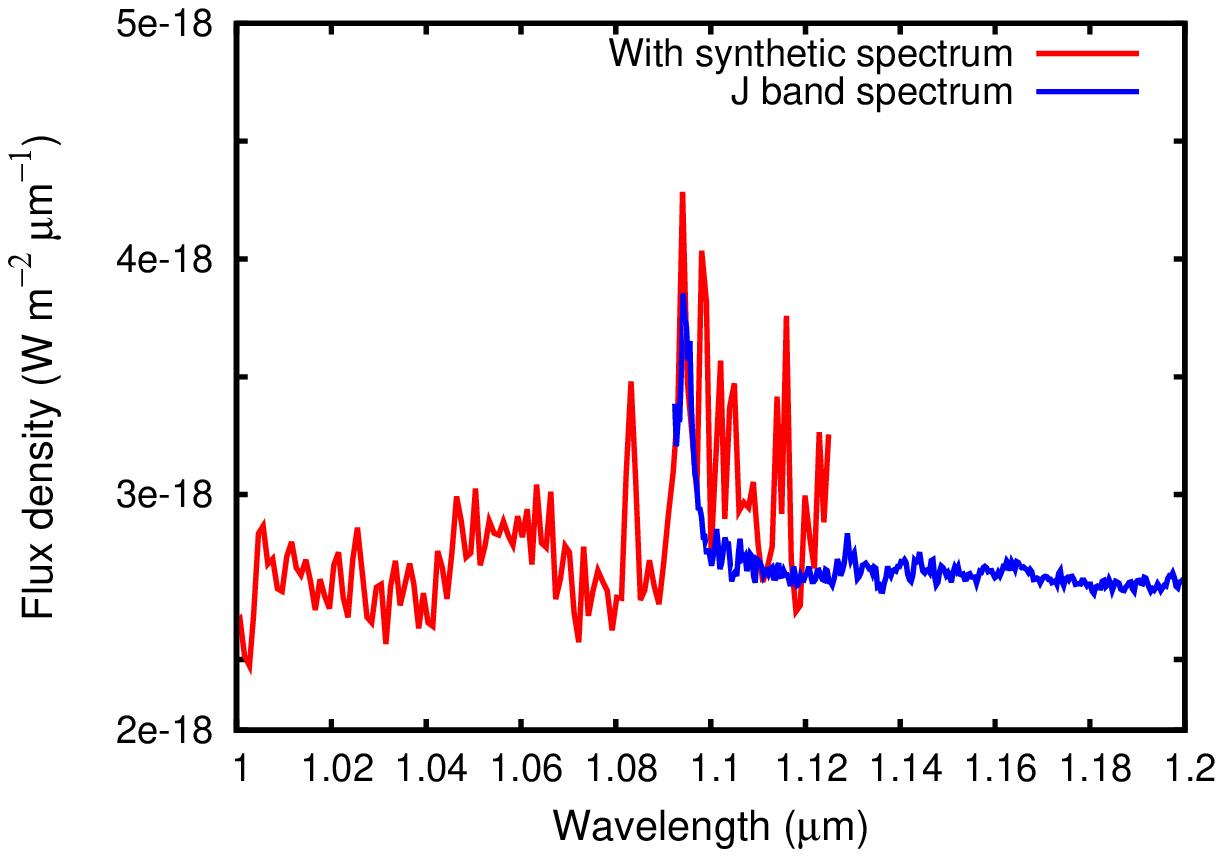}\\
\includegraphics[width=9cm]{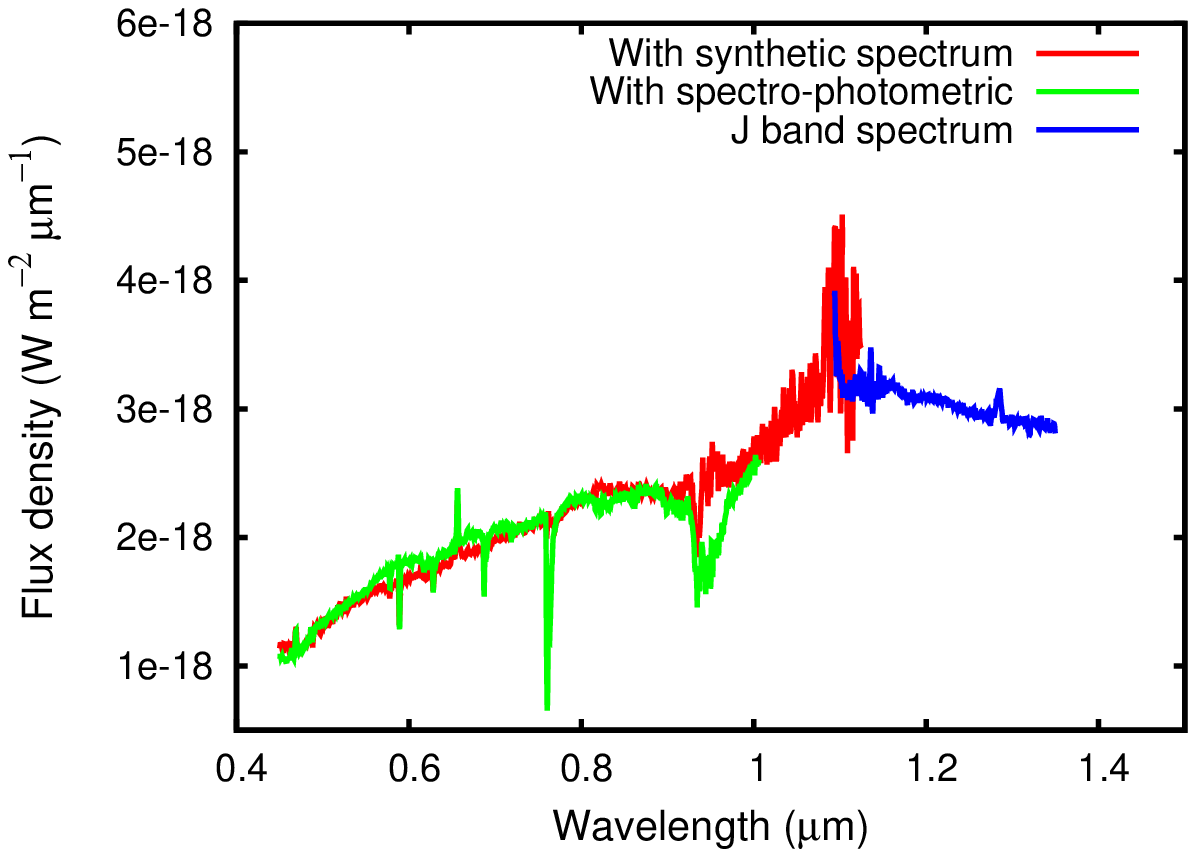}&\includegraphics[width=9cm]{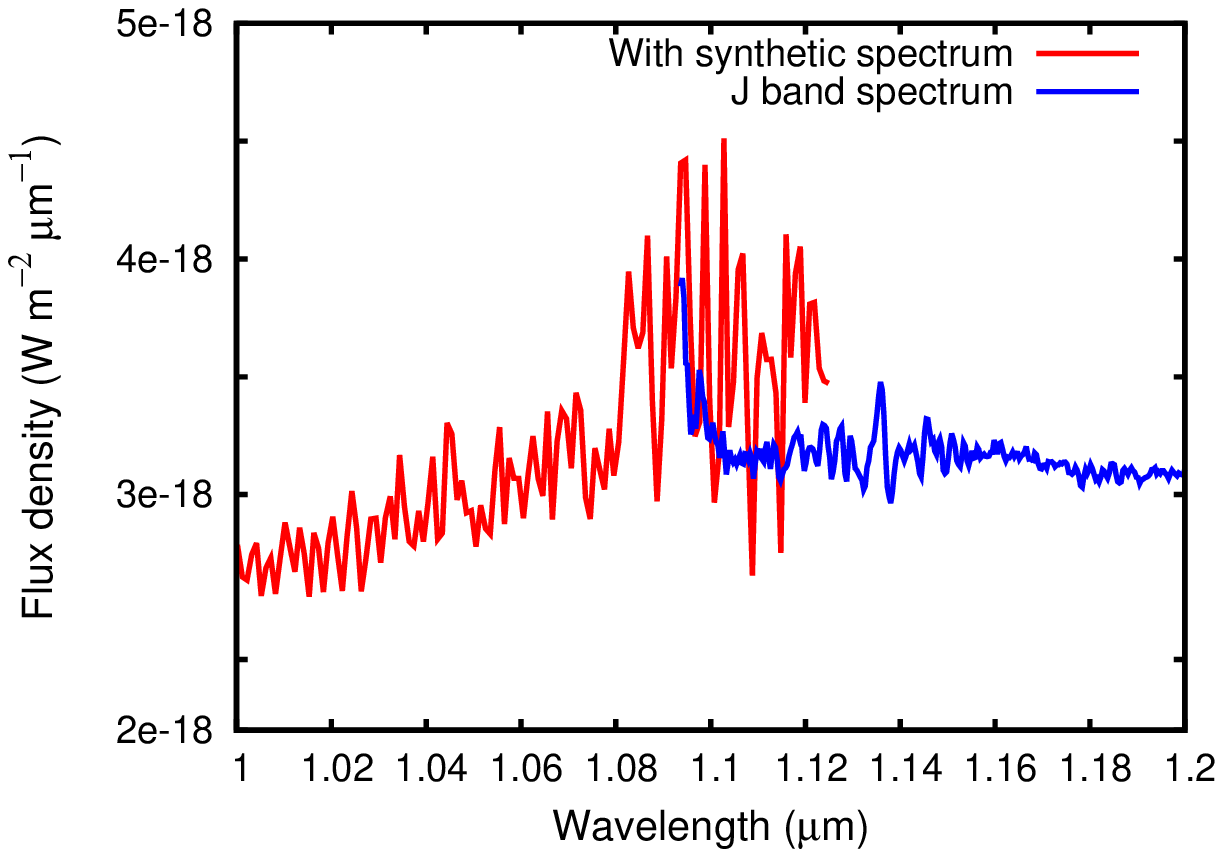}\\
\includegraphics[width=9cm]{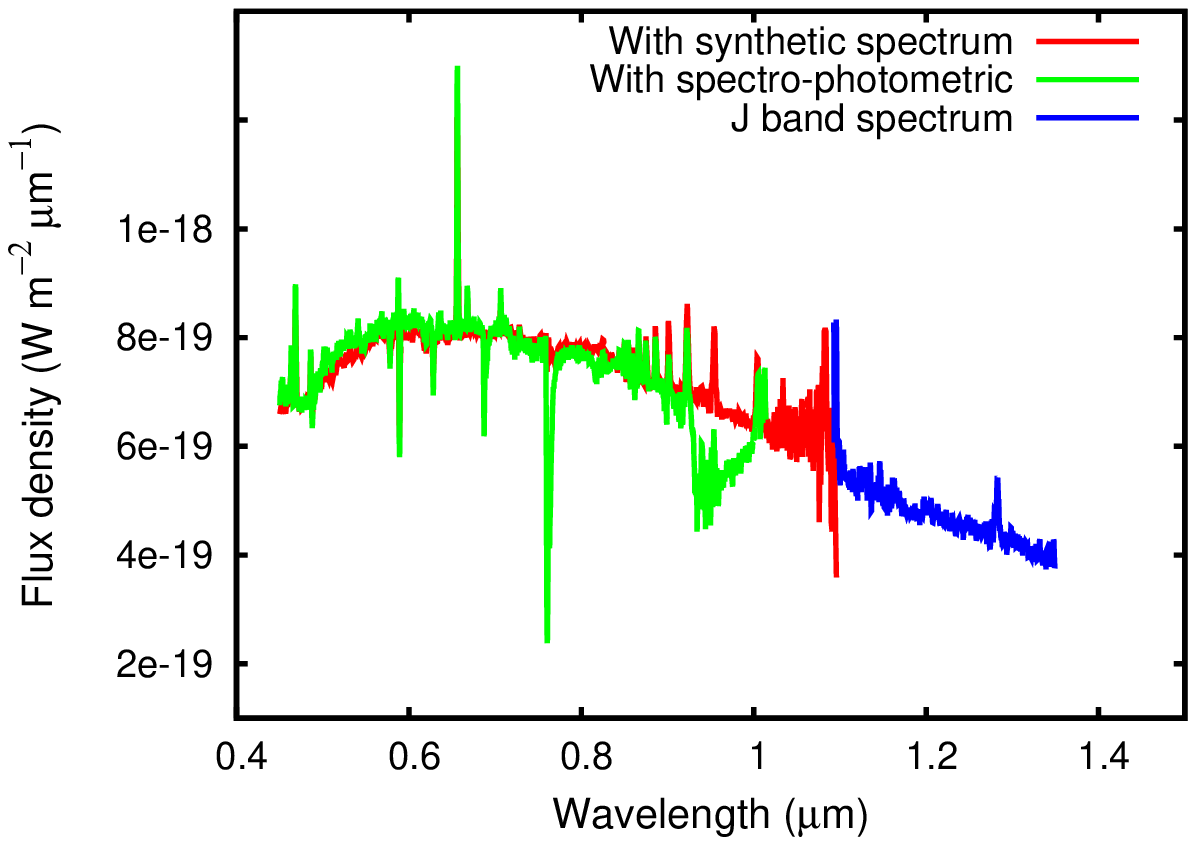}&\includegraphics[width=9cm]{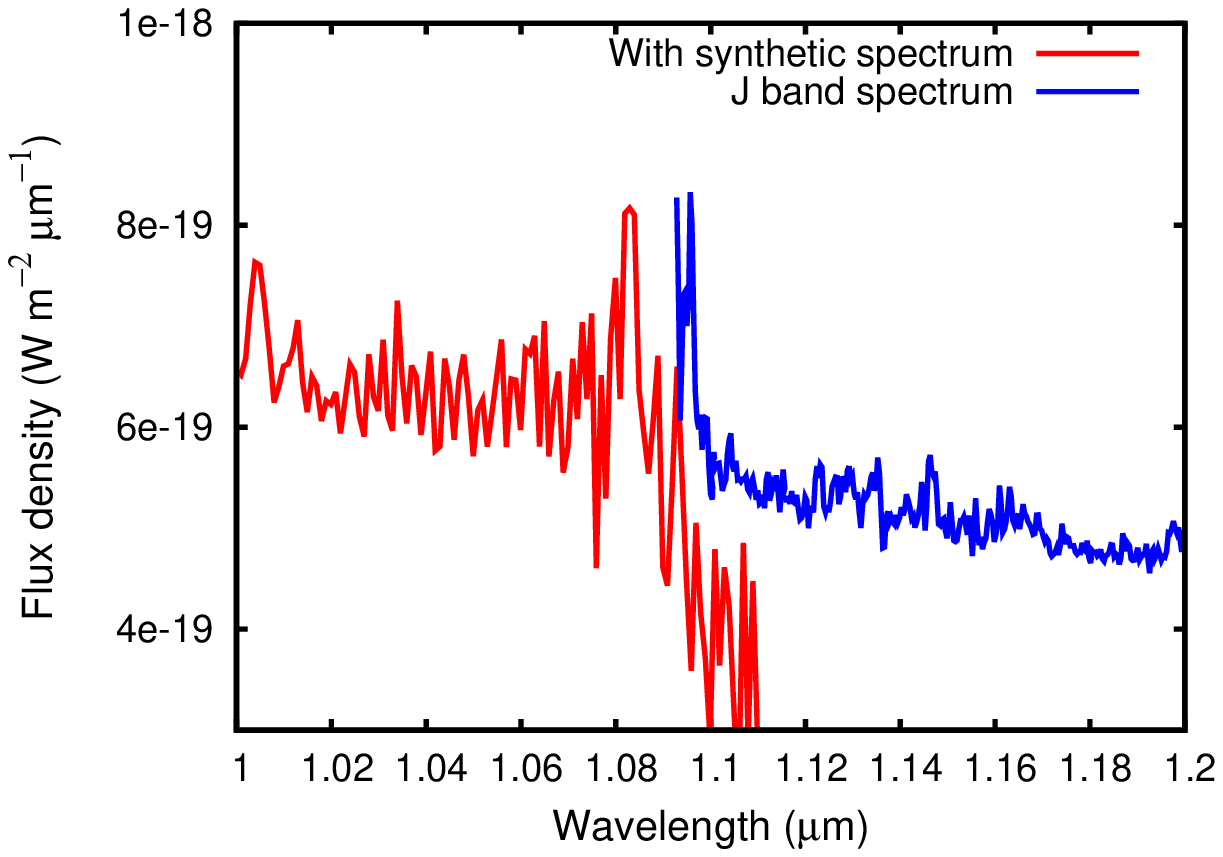}\\
\end{tabular}
\caption{\small Observed optical and {\it J} band spectra of \gx\
  during Obs.~1 (top), 2 (middle), and 3 (bottom), full-scale (left)
  and zoom-in on the overlapping region (right). The optical
  spectra derived using a synthetic standard star template are
  displayed in red and the ones using the ESO  spectro-photometric
  calibration in green. The {\it J} band spectra are displayed in
  blue. Note that the telluric absorption features are still present
  in the green spectra. The red spectra displayed here are rebinned to a third of
  their real resolution to improve the visibility of the optical/\textit{J} overlapping
  region. However, only the original ones were used for SED fitting.}
\label{compopt}
\end{center}
\end{figure*}

Figure~\ref{compopt} displays a comparison, for each observation,
between the optical spectra of \gx\ obtained using either the ESO spectro-photometric
calibration or Kurucz templates (left panel), as well as a zoom-in the
optical/\textit{J} overlapping region (right panel). The use of synthetic spectra is 
necessary to extend the wavelength coverage and cover the gap with the
near-IR.
\clearpage

%\bsp

\label{lastpage}

\end{document}